\documentclass[12pt]{article}
\usepackage{times}
\usepackage{subfig}
\usepackage{geometry}
\usepackage[utf8]{inputenc}
\usepackage{enumitem,amssymb}
\usepackage{float}
\usepackage{longtable}
\usepackage[toc,page]{appendix}
\usepackage[table,xcdraw]{xcolor}
\usepackage{url}
\usepackage{hyperref}
\usepackage[export]{adjustbox}
\usepackage{mathptmx}
\usepackage{booktabs,caption}
\usepackage{amsmath}
\usepackage[utf8]{inputenc}
\usepackage[T1]{fontenc}
\usepackage{dcolumn,booktabs,rotating}
\usepackage{cite}
\usepackage{ragged2e}
\usepackage[table,xcdraw]{xcolor}
\usepackage{lipsum}
\usepackage{setspace}
\usepackage{pdfpages}
\usepackage{ltablex}
\usepackage{booktabs,tabularx,ragged2e,dcolumn,caption} 
\usepackage[colorinlistoftodos]{todonotes}
\usepackage{array}
\newcolumntype{C}[1]{>{\centering\arraybackslash}p{#1}}

\usepackage{booktabs,caption}
\usepackage[flushleft]{threeparttable}
\usepackage{dcolumn}
\usepackage{booktabs}
\usepackage{rotating}
\usepackage{tabularx}
\usepackage{ragged2e}
\usepackage{lscape} 
\usepackage{lmodern}
\usepackage{microtype}
\usepackage{float}
\usepackage{caption}
\usepackage{booktabs}
\usepackage{blindtext}
\usepackage{siunitx}
\usepackage{csvsimple}
\usepackage{pdfpages}
\usepackage{amssymb}
\usepackage{array}
\usepackage{authblk}
\usepackage{multirow}
\usepackage{makecell}
\usepackage{graphicx}
\usepackage{booktabs}
\usepackage{pifont}

\newcommand{\distas}[1]{\mathbin{\overset{#1}{\kern\z@\sim}}}%
\newsavebox{\mybox}\newsavebox{\mysim}
\newcommand{\distras}[1]{%
  \savebox{\mybox}{\hbox{\kern3pt$\scriptstyle#1$\kern3pt}}%
  \savebox{\mysim}{\hbox{$\sim$}}%
  \mathbin{\overset{#1}{\kern\z@\resizebox{\wd\mybox}{\ht\mysim}{$\sim$}}}%
}
\makeatother

\title{Product recalls, market size and innovation in the pharmaceutical industry}
\author[1]{Andrea Morescalchi\thanks{andrea.morescalchi@imtlucca.it}}
\author[2]{Federico Nutarelli\thanks{federico.nutarelli@imtlucca.it}}
\author[3]{Massimo Riccaboni\thanks{massimo.riccaboni@imtlucca.it}}
\affil[1]{AXES Group}

\date{IMT School for Advanced Studies}

\begin{document}
\maketitle
\begin{abstract}

\noindent The idea that research investments respond to market rewards is well established in the literature on markets for innovation (Schmookler, 1966; Acemoglu \& Linn, 2004; Bryan \& Williams, 2021). Empirical evidence tells us that a change in market size, such as the one measured by demographical shifts, is associated with an increase in the number of new drugs available (Acemoglu \& Linn, 2004; Dubois et al., 2015). However, the debate about potential reverse causality is still open (Cerda et al., 2007). In this paper we analyze market size's effect on innovation as measured by active clinical trials. The idea is to exploit product recalls an innovative instrument tested to be sharp, strong, and unexpected. The work analyses the relationship between US market size and innovation at ATC-3 level through an original dataset and the two-step IV methodology proposed by Wooldridge et al. (2019). The results reveal a robust and significantly positive response of number of active trials to market size.

\end{abstract}

\section{Introduction}
Exploring the actual relationship between market rewards and innovation has been widely explored in innovation economics for a long time (\cite{scherer1982demand}, \cite{schmookler2013invention}, \cite{klepper2010demand}). This opened to the possibility of public demand in stimulating innovation, such as in the case of orphan drugs. Schmookler's "demand-pull" hypothesis, implying that innovation is a function of market demand, has been challenged over the years. Already in the '90s \cite{kleinknecht1990demand} noticed that the direction of causality between market size and innovation appears to be far from obvious. In particular, the authors suggested the presence of a simultaneous relationship between demand and innovation but did not manage to control for it. More recently, \cite{stoneman2010soft} and \cite{ball2018product} developed more rigorous ways to detect such type of endogeneity.\\
Acemoglu and Linn (2004) developed a strategy to overcome the endogeneity bias at the market level. Specifically, they exploited changes in the market size for different drug categories driven by U.S. demographic trends (\cite{acemoglu2004market}). After the contribution of \cite{acemoglu2004market}, the focus moved from ascertaining the presence of the reverse causality of market size and innovation to detecting the best instrument for market size. Indeed, the instrument adopted in \cite{acemoglu2004market} was later criticized by \cite{cerda2007endogenous} as being itself endogenous. As detailed in \cite{cerda2007endogenous}, while pharmaceutical innovation increases the age of patients, the fact that the average age increases imply that more patients would need innovative products. To the best of our knowledge, such a gap in the literature is still unfulfilled.  \\
Besides, authors, pushed by the studies of \cite{acemoglu2004market}, mainly concentrate their efforts on the Pharmaceutical industry. The Pharmaceutical, indeed, constitutes a definitive case study: in such a sector, consumers' needs are diverse and almost constant over time, which allows to separate it into independent sub-markets based on such needs  (\cite{bertoni2010increased}).  Furthermore, investments in innovation are vital for the industry's existence. Innovation is also easily measurable. In the Pharmaceutical industry, market size is defined based on the Anatomical Therapeutic Chemical (ATC) Classification System, i.e., a drug classification system that classifies the active ingredients of drugs according to the organ or system on which they act and their therapeutic, pharmacological, and chemical properties. \\
The present paper aims at capturing the relationships among market size and innovation at the ATC-3 level by instrumenting market size with recalls (see below) of drugs operated by the Food and Drug Administration (FDA). Our work tries to contribute to the literature in three ways. \\
First, we adopt an innovative measure of innovation, i.e., the overall number of trials at the ATC-3 level instead of the cumulative R\&D expenditures or New Molecular Entities (NME). This necessary modification overcomes the limitations of the other two most adopted measures. R\&D expenditures are, indeed, linked both to firms' long-term profit decisions (Cohen, 2010) and, more critically, to their size. \cite{stoneman2010soft}, among others, suggested that smaller entrants might be more inclined to invest in R\&D expenditures than their bigger veteran competitors. Hence, innovation as cumulative R\&D expenditures of the firms composing the market might be related to market size as measured by the firms' cumulative sales composing the market. More delicate is the topic concerning NME, also adopted in \cite{acemoglu2004market}. NME are innovative products containing active moieties that have not been approved by the FDA previously, either as a single ingredient drug or as part of a combination product. They can be either innovative new products that never have been used in clinical practice or the same as, or related to, previously approved products. Though a complete definition, the one of NME does not fully capture, in our opinion, the will of innovation by firms inside the market. The reason hinges around the stage of drug approvals at which NME, with respect to the measure of innovation employed in the present work, are approved by FDA.\\
Pharmaceutical drug approval is a long process. Firms should first pass a pre-clinical phase, a stage of research that starts before that clinical trials (testing in humans) can begin, and during which important feasibility, iterative testing, and drug safety data are collected. The clinical drug development stage, then, consists of three phases. In Phase 1, clinical trials are conducted using healthy individuals to determine the drug's basic properties and safety profile in humans. Typically, the drug remains in this stage for one to two years (\cite{dimasi2003price}). In Phase 2, efficacy trials begin as the drug is administered to volunteers of the target population. Finally, Phase 3 compares a new drug to a standard-of-care drug. NME are FDA-approved entities having overcome pre-clinical trials, while the number of trials also considers the pre-clinical phase. In other words, the latter measure also considers potentially unsuccessful trials, i.e., trials not passed to the clinical phase, which equally characterize an innovative drive of the firm. Furthermore, the number of trials, differently from NME, includes the definition of innovation and the drugs that the FDA did not approve for clinical trials.\\ 
A second contribution is given by the adoption of more refined ATC classes. The available data on the ATC-3 level of classification well captures the structure of sub-markets, usually constructed artificially or disregarded by literature. Moreover, the ATC-3 level is the level employed by antitrust agencies. Refer to Section \ref{data} for details.\\
A further improvement is methodological. The paper adopts an IV approach to deal with the endogeneity problem of market size. The enhancement compared to past research consists of the instrumentation of market size with recalls to overcome the endogeneity issue already detailed. The idea is to exploit sharp and unexpected recalls. The task of characterizing recalls as being sharp and unexpected requires a general definition of recalls. We believe that recalls are exogenous (as detailed in Section 3).\\
FDA refers to a recall as the most effective way to protect the public from a defective or potentially harmful product. A recall is a voluntary action taken by a company to remove a defective drug product from the market. Drug recalls are conducted either on a company's initiative or by FDA request. In a recall, the FDA's role is to oversee a company's strategy, assess the recall's adequacy, and classify the recall. According to their severity, the FDA classifies the recalls in Class I  (more severe), Class II, and Class III (least severe). Medicines may be recalled for several reasons ranging from health hazards to potential contamination, adverse reaction, mislabeling, and poor manufacturing. Recalls should not be confused with withdrawals. Unlike the FDA definition, literature often refers to withdrawals as post-marketing recalls imposed by the FDA on firms due to their high severity and risk to human health. Therefore, recalls can be either expected and voluntarily made by firms if minor or sharp and unexpected if most severe and forced by the FDA.\\
By the very definition of drug recalls, we expect a drop in sales consequent to a drug recall in a market. To clarify the latter mechanism, one can refer to Merck's popular recall of VIOXX in 2004. VIOXX was withdrawn from the market due to an increased risk for serious cardiovascular events. The recall caught unprepared both the market and the firm. After the announcement of the recall of VIOXX in September 2004, shares of Merck and its sales dropped. This drop has been publicized by mass media (\cite{nyt}, \cite{fit} among others) and well recognized by academics (see, e.g. \cite{tong2009vioxx} among others).\\
The present work tries to assess such sharp and unexpected recalls (" major recalls" from now on). The definition of major recalls we adopted throughout the paper has been recovered by filtering the causes of Class I recalls. We filtered recalls according to the relevance of the cause, its severity in terms of potential danger against human life, and the FDA's actions. Specifically, we comprised in the definition of major recalls, withdraws, Class I recalls containing critical keywords among their causes such that:" contamination," "death/s," "overdose," "symptoms," "particulate matters," and "adverse reaction." \\
We did not employ Class II recalls since we believe that they constitute a weaker instrument than major recalls. Nonetheless, we included in the Appendix the analysis using all types of recalls as a robustness check. The main results are confirmed.\\
To the best of our knowledge, it is the first time that recalls are employed as an instrument for market size. 
\section{Literature Review} \label{lit_rev}
The literature acknowledges the importance of market size in explaining the rate of innovation for many years. Back in 1942, Schumpeter
indicated that larger firms are more innovative than smaller ones.
In the early '60s, the focus shifted more broadly on the possible effects of demand on market size (see, e.g., \cite{scherer1982demand}). It was not yet clear whether the reverse causality of demand and innovation played a relevant role. \cite{scherer1982demand}, for instance, argued that causality ran primarily from sales to innovation. His study, however, has been criticized in several aspects. The definition of demand was indeed still too broad and was not conclusive about the unique sign on the relationship between demand and innovation (see, e.g., \cite{mowery1979influence}). At the time, the research did not focus specifically on the pharmaceutical sector nor looked at the aggregate market level (see, e.g., \cite{pakes1984rate}).\\
Most recently, \cite{kleinknecht1990demand} denounced a clear reverse causality of demand and innovation, thus invalidating the prior studies. \cite{geroski1995innovative} empirically verified such conclusions soon after, finding out how innovations increase demand by creating their demand.\\
Besides, it was clear that heterogeneous shifts of demand played a prominent role in determining technological development (see, e.g., \cite{malerba2007innovation}). Between 1980 and 1990 and most recently in 2002, several studies showed, for instance, how innovation reacted elastically to energy prices.\\ 
Nowadays, a huge part of the research on the relationship between market size and innovation regards the pharmaceutical industry, where innovation represents a pushing power. Literature mainly takes into account two levels of aggregation: firm-level and market-level. Past research efforts have been devoted to identifying the impact of
firm size on R\&D investments and output. Nevertheless, this
 question is still an open debate ( see \cite{mellahi2010study}, \cite{kolluru2017empirical} among others). Specifically, controversial results emerge due to the difficulty in fully excluding unobservable endogeneity sources varying with time.  Such unobservables might derive from strategic decisions taken within the firms, which, in turn, might be related to their size. For example, small pharmaceutical firms are likely to take more risky decisions than big established ones (\cite{hall2010handbook}). Moving to market aggregation easily avoids the mentioned concerns. Unobservables related to market size can principally be considered as intrinsic characteristics of markets and, consequently, fixed in time. Thus, fixed effect techniques allow researchers to control for unobservable heterogeneity, purging the idiosyncratic endogeneity of market size. Therefore, the market seemed a more suitable level, and most authors shifted to the latter level of aggregation. \\
The literature of the pharmaceutical sector is varied. Part of its variability is due to the measures of innovation adopted. Some authors adopted accounting data focusing on R\&D. The latter, though robust under perfect capital markets 
becomes inconclusive with imperfect markets.  Because current revenues (market size) are a reasonable proxy for future market size, and since present R\&D may be responding both to present and to future sales opportunities, results incorporate two effects that are difficult to separate. Aware of such an issue, the authors included lagged proxies of the market size (see, e.g., \cite{giaccotto2005drug} who estimated that a 1\% increase in price leads to a 0.58\% increase in R\&D spending).\\
Other measures of innovation include clinical trials (see \cite{kyle2012investments} among others) and changes in Medicare part D affecting both present and future market size (\cite{blume2013market}). Scholars found a positive response of innovation to shocks in market size. Again, the problem remained the possible co-occurrence in innovation's response to both current and expected cash flows generated by market size shocks.\\
Besides, innovation has been quantified by the number of relevant journal articles about a condition (\cite{lichtenberg2006pharmaceutical}). Further measurements comprise the number of new drugs launched, including generic drugs (\cite{acemoglu2004market}, \cite{dubois2015market}) in the form of new molecular entities (NME), new chemical entities (NCE), or approvals of new medicines by the FDA. \\
Similarly, many measures of market size have been embraced. \\
\cite{acemoglu2004market}  gave a first significant contribution on the relation between market size and innovation in the pharmaceutical industry. Their idea relies on adopting demographic shifts to instrument market size and control for the endogeneity arising from reverse causality. In particular,  \cite{acemoglu2004market} exploit variations
in the expenditure share of different U.S. age cohorts for different therapeutic classes from 1970-2000. They find that a 1\% increase in expenditure shares leads to a 4\% increase in the number of new drugs, a far higher elasticity than the average elasticity found in the remaining literature (\cite{dubois2015market}). 
\cite{cerda2007endogenous} provided further insights on the results found in \cite{acemoglu2004market}. Employing U.S. demographic data, \cite{cerda2007endogenous} showed that there are essential feedback effects not considered in \cite{acemoglu2004market}. New drugs might affect the market size through their impact on the mortality rate. Indeed, innovative medicines are likely to cure more diseases, raising the population's average age and, hence, the number of older people needing such cures. Demand shifts accordingly, bringing out again the issue of reverse causality.\\
Recent literature on the topic improves above all on the methodological part (see e.g. \cite{lichtenberg2006pharmaceutical}, \cite{civan2009effect}, \cite{dubois2015market}, \cite{rake2017determinants} and others). Authors found, on average, that a 1\% increase in the market size measure increases innovation of 0.4\% to 0.7\%.\\
Past papers acted mainly at disease level or, at most,  at ATC-1 or ATC-2 levels (see e.g.\cite{dubois2015market}). To the best of our knowledge, no works are focusing on the more interesting ATC-3 level at which antitrust authorities work.
According to us, there are several advantages of using drug classes rather than disease classes. Firstly, since firms request NCTs, NMEs, and NDAs directly, devoting too much attention to the demand-side might neglect the supply-side dynamics, which induce firms to undergo an NDA. In particular, aggregate sales of drugs align to the supply-side, while sales based on disease classes (i.e., aggregated sales of products purchased by patients) are more on the demand side. \\
In other words,  while firms might follow demand-side stimuli to undergo an NDA (or an NCT), they might also look up at the competitors, i.e., products of other companies in the same ATC class. The latter applies to commercial trials when the sponsor is a pharmaceutical industry and not academy/research related. \\
Secondly, by taking disease classes, one includes in the definition of innovation different chemical and therapeutic typologies of drugs ranging from topical to systemic drugs, from vaccines to ointment. This lack of distinction might lead to endogeneity through several channels, such as people's expectations. Patients might beware of some drugs, affecting the probability of having a larger market size for the product's typology under question. Other endogeneity sources regard the possibility of a correlation between regressors and the error term (which includes "drug-type").  For instance, regulations may be product type-specific (e.g., the regulation of the WHO vaccines do not apply to other drug types). Other possibly problematic controls are knowledge stocks, which could again depend on the product type.  Moreover, knowledge stocks might increase by developing innovative medicines in classes where only a particular type of medicine has been developed until that moment.  An example is provided in dermatology, where academics produce papers for adopting topical medicines for systemic usage due to some systemic medicines' undesired side effects. \\
Finally, the length of a clinical trial varies depending on the type of medicine under study, which may cause lagged effects of market size if disease class is employed. \\
To the best of our knowledge, among the several innovation measures, no work exploited INDs and early stages clinical trials (i.e., pre-clinical and Phase I) together with Phase II and Phase III trials. \\
The two more recent estimates of the relationship between market size and innovation have been provided in \cite{rake2017determinants} and \cite{dubois2015market}. The latter used NCE to measure innovation and defined market size as a measure of expected revenue. The dataset comprised information about sales for 14 different countries. Specifically, \cite{dubois2015market} measured market size as the total revenue over the entire life cycle of a branded drug. \cite{dubois2015market}  performed a control function approach and recovered an estimate of the relation between market size and innovation for each therapeutic class at level 1. The average elasticity of innovation to the market size in \cite{dubois2015market} was about 23\%, which is relatively low than the average estimates. A possible explanation can be found in \cite{blume2013market}, which states that several of the countries chosen for the analysis regulate prescription drug prices, and regulations may change rapidly over time. Thus, given the lower expected profit per consumer and more significant uncertainty about future profits and prices, firms' R\&D decisions are likely to be less responsive to a unit change in expected revenues for all these countries combined versus the exact unit change in the U.S. market.\\
Finally, \cite{rake2017determinants} adopted several measures of innovation from NCE to clinical trials in Phase II and Phase III. \cite{rake2017determinants} found no evidence of reverse causality when adopting NCE. One of his efforts was to account for the fact that changes in the industry's R\&D process, from "random screening" to "guided drug development," pointing out the importance of advances in molecular biology and related fields (\cite{rake2017determinants}). The author modeled technological opportunities and inserted them as a regressor in the analysis, finding a positive relationship with Phases II and III trials. His results are in line with \cite{cerda2007endogenous} and \cite{acemoglu2004market}. \\


Tab. \ref{litrev} in Appendix provides a schematic literature review on previous estimates of the relation between innovation and market size.

\section{Data}\label{data}
\begin{figure}[H]%
    \centering
    \subfloat[\centering \footnotesize Number of trials and INDs per year]{{\includegraphics[width=7.7cm]{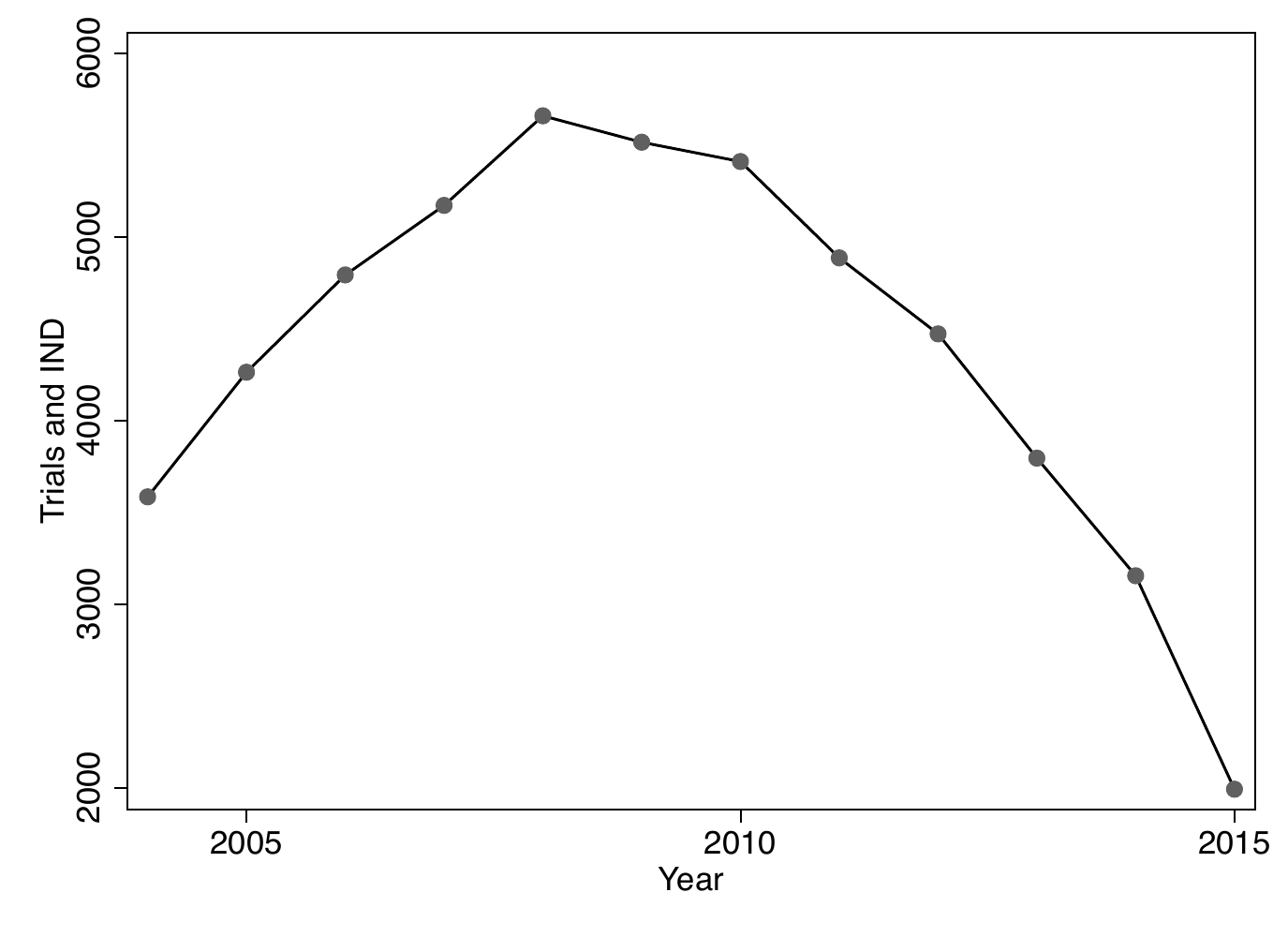} }}%
    \qquad
    \subfloat[\centering Trend of sales by year]{{\includegraphics[width=7.7cm]{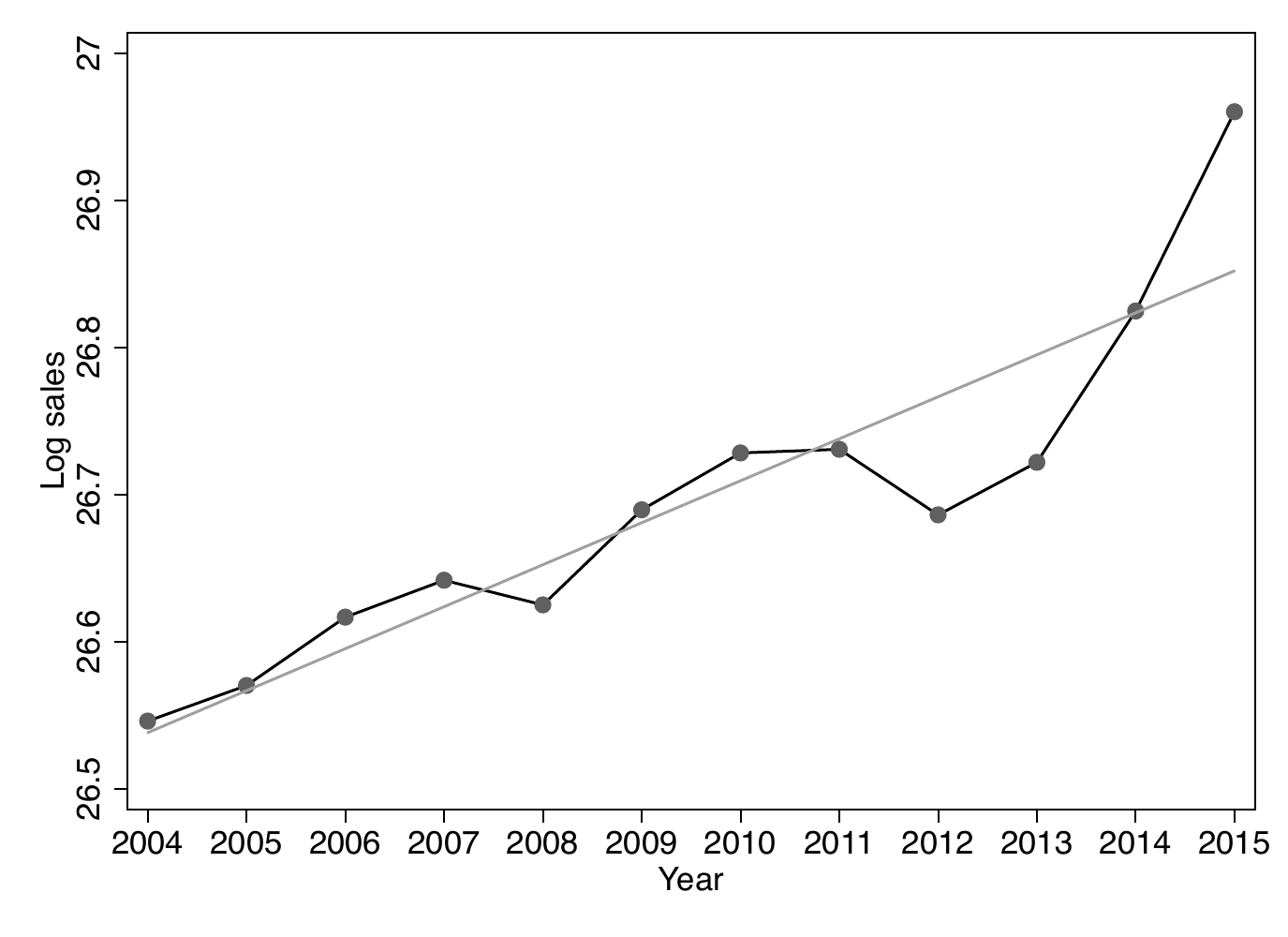} }}%
    \caption{Overview of sales' and trials' trends}%
    \label{tr_yearly}%
\end{figure}

The sales data employed come from Evaluate dataset. The controls have been extrapolated from Evaluate, from the PHarmaceutical Industry Database (PHID) and FDA. Specifically, some of the regressors derive from an elaboration of the variables present in the PHID database.\\
Sales data for the US pharmaceutical market ranges from 2004 to 2015. Sales data were initially available at the product and molecule level and have successively been aggregated at the ATC-3 level. 
In the ATC classification system, drugs are classified at five levels (ATC-1, ATC-2, ATC-3, ATC-4, ATC-5): the higher the level, the more detailed the classification. 
Acemoglu, Linn (2004) employed ATC-1 and ATC-2 categorizations to define market size. In particular, Acemoglu, Linn (2004) constructed market size as the sum of the average expenditure share of drugs in an ATC-1 (ATC-2) category across all ATC-1 (ATC-2) categories. \\
Data at our disposal allow us to catch the diverse strata of products inside broader classes (ATC-1 and ATC-2) in terms of both demand and supply dynamics. Medicines classified inside an ATC-1 or an ATC-2 level can satisfy patients with completely diverse needs since they are designed to cure various diseases. At the same time, a firm investing in the same ATC-2 sector might invest in more ATC-3 sectors. In the case of ATC-1 or ATC-2 adoption, the latter missing information may lead to the construction of uninformative innovation and market size variables.  Such controls might not consider the firms' specialization in a sub-sector rather than in another one belonging to the same ATC-2 or ATC-1 class. \\ In previous work, we have also evaluated other levels of analyses (firm, product, and ATC-firm aggregations) (\cite{nutarelli}) but opted for ATC-3 level because of the importance of ATC-3 level being employed by antitrust agencies. To provide some examples, we mention  Provost et al.(2019), Markham, A. (2020), Vaishnav, A. (2011), Hawk et al.(2000), Cheng J. (2008), and other cases mostly pertaining M\&A (e.g., Case M.8889 - TEVA / PGT OTC ASSETS of 2018). \\
We avoided adopting the ATC-4 level since, at such granularity, products belonging to a specific ATC-4 class might not differ substantially from others belonging to another ATC-4 class. This might lead to between-group dependencies (e.g., innovations in an ATC-4 at level 4 may also affect a close ATC-4 class) which could cause inference to be invalid. Further, at the ATC-4 level, compensations may also intervene between groups, thus invalidating the strength of the instrumental variable recalls. \\ 
The available data also contain the launch date and ATC code of products. We focused on worldwide sales of US companies. \\
Data on New Clinical Trials (NCT) for 2004-2015 at product level come from the ClinicalTrials.gov website, while data on commercial Investigational New Drugs (IND) at product level derive from a Pharmaceutical Industry Database maintained at IMT Lucca. \\
Clinical trials are research studies performed on people who aim to evaluate a medical, surgical, or behavioral intervention. An IND in clinical trials is the mean by which a pharmaceutical company obtains permission to start human clinical trials and to ship an experimental drug across state lines before a marketing application for the drug has been approved.\\
Clinical trials comprise trials from Phase I to Phase IV. Fig.\ref{tr_yearly} displays the yearly number of trials and commercial IND as obtained by the mentioned sources.\\
It also shows the expected positive trend of sales of the Pharmaceutical industry in time.

A considerable drop in Trials and IND occurred after 2013, as it is evident from Fig.\ref{tr_yearly} (a). The reason for such lack is that, in general, clinical trials innovate drugs, approaches, and interventions. However, approaches and interventions are excluded from the count of trials to focus strictly on innovation coming from industrial sources.\\

Recalls data have been manually collected from different sources, among which FDA website, openFDA, various articles, and web sources (e.g., \cite{onakpoya2016worldwide}; WHOCC website, PubMed, \cite{siramshetty2016withdrawn} and others).\\
The 7.19\% of the firms' sample (i.e., 697 firms in total) have issued Class II recall. Among the firms that issued a recall, 51 firms underwent a recall of Class I, 27 of which issued a single recall of Class I, and just three firms issued more than 9 Class I recalls.\\
The provided estimates must be read in light of the database's possible limitations regarding the presence/absence of firms and products inside it. \\
Besides, the recalls of pure compounders were only partially included \footnote{Due to the unavailability of data. We verified that the representativeness of recalls is preserved \cite{nutarelli}} and, when included, were attributed to the unique manufacturer/distributor in the database. Finally, recalls coming from repackaging firms were not attributed uniquely to the repackager (e.g., Aidapak) but the labeler specified in the NDC. \\
Due to these case-specific engines, it is not easy to establish a  unique and unambiguous pattern of recalls over the years. The situation is further complicated wherever different resources employ different methodologies to count the recalls. An example of the cited uncertainty in sources is found when comparing \cite{duemilanovequindici} and \cite{laguna}. Specifically, \cite{duemilanovequindici} asserts that the number of recalled products had remained reasonably constant except for 2010 and 2013 when the number went down by approximately 35\%. The statement contrasts with what was reported in \cite{laguna}. According to \cite{laguna}  \textit{"a spike in the number of drugs recalled occurred in 2013. There were nearly 60 recalls in that year alone. However, 2017, with 71 recalls, saw nearly the highest number of recalls since 2009. Only 2011 and 2009 surpassed it at 74, and 75 recalls, respectively"}.\\
Tab. \ref{refere}, provides a list of the primary sources and the average number of recalls across them. The following is an attempt to overcome the mentioned issues involving the dissimilarities of data origins.

\begin{table}[H]
\scalebox{1}{
\begin{tabular}{llllll
>{\columncolor[HTML]{EFEFEF}}l }
\hline
\multicolumn{1}{|l|}{\cellcolor[HTML]{C0C0C0}{\color[HTML]{000000} \textbf{Year}}} & \multicolumn{1}{l|}{\cellcolor[HTML]{C0C0C0}{\color[HTML]{000000} \textbf{CNN}}} & \multicolumn{1}{l|}{\cellcolor[HTML]{C0C0C0}{\color[HTML]{000000} \textbf{Regulatory Focus}}} & \multicolumn{1}{l|}{\cellcolor[HTML]{C0C0C0}{\color[HTML]{000000} \textbf{\cite{hall2016characteristics}}}} & \multicolumn{1}{l|}{\cellcolor[HTML]{C0C0C0}{\color[HTML]{000000} \textbf{FDA Enforcement Reports}}} & \multicolumn{1}{l|}{\cellcolor[HTML]{C0C0C0}\textbf{\cite{laguna}}} & \multicolumn{1}{l|}{\cellcolor[HTML]{C0C0C0}{\color[HTML]{000000} \textbf{AVERAGE}}} \\ \hline
\textbf{2004}                                                                      &                                                                                  & 68                                                                                            &                                                                                                                                       &                                                                                                      &                                                                                 & 68                                                                                   \\
\textbf{2005}                                                                      &                                                                                  & 140                                                                                           &                                                                                                                                       &                                                                                                      &                                                                                 & 140                                                                                  \\
\textbf{2006}                                                                      & 384                                                                              & 109                                                                                           &                                                                                                                                       &                                                                                                      &                                                                                 & 243                                                                                  \\
\textbf{2007}                                                                      & 391                                                                              & 56                                                                                            &                                                                                                                                       &                                                                                                      &                                                                                 & 189                                                                                  \\
\textbf{2008}                                                                      & 426                                                                              & 128                                                                                           &                                                                                                                                       & 176                                                                                                  &                                                                                 & 244                                                                                  \\
\textbf{2009}                                                                      & 1742                                                                             & 85                                                                                            &                                                                                                                                       & 1660                                                                                                 &                                                                                 & 890                                                                                  \\
\textbf{2010}                                                                      &                                                                                  & 135                                                                                           &                                                                                                                                       & 389                                                                                                  &                                                                                 & 262                                                                                  \\
\textbf{2011}                                                                      &                                                                                  & 236                                                                                           &                                                                                                                                       & 1279                                                                                                 & 75                                                                              & 530                                                                                  \\
\textbf{2012}                                                                      &                                                                                  & 381                                                                                           & 499                                                                                                                                   & 1518                                                                                                 &                                                                                 & 799                                                                                  \\
\textbf{2013}                                                                      &                                                                                  & 1031                                                                                          & 1283                                                                                                                                  & 848                                                                                                  & 60                                                                              & 805                                                                                  \\
\textbf{2014}                                                                      &                                                                                  & 640                                                                                           & 1344                                                                                                                                  & 893                                                                                                  &                                                                                 & 959                                                                                  \\
\textbf{2015}                                                                      &                                                                                  &                                                                                               &                                                                                                                                       & 1584                                                                                                 &                                                                                 & 1584                                                                                
\end{tabular}}
\caption{Sources with reported number of recalls}
\label{refere}
\end{table}

To overcome the dissimilarities of data origins, we chose the average as the benchmark to compare with the collected recalls. Fig.\ref{number_recalls_our_vs_b} illustrates in more detail the comparison between the benchmark recalls represented by the average recalls among all sources and the collected recalls. Aidapak's recalls of 2011 are considered as "outliers" and, for this reason, are not included among the collected recalls at this stage:

\begin{figure}[H]
\centering
  \includegraphics[width=.8\linewidth]{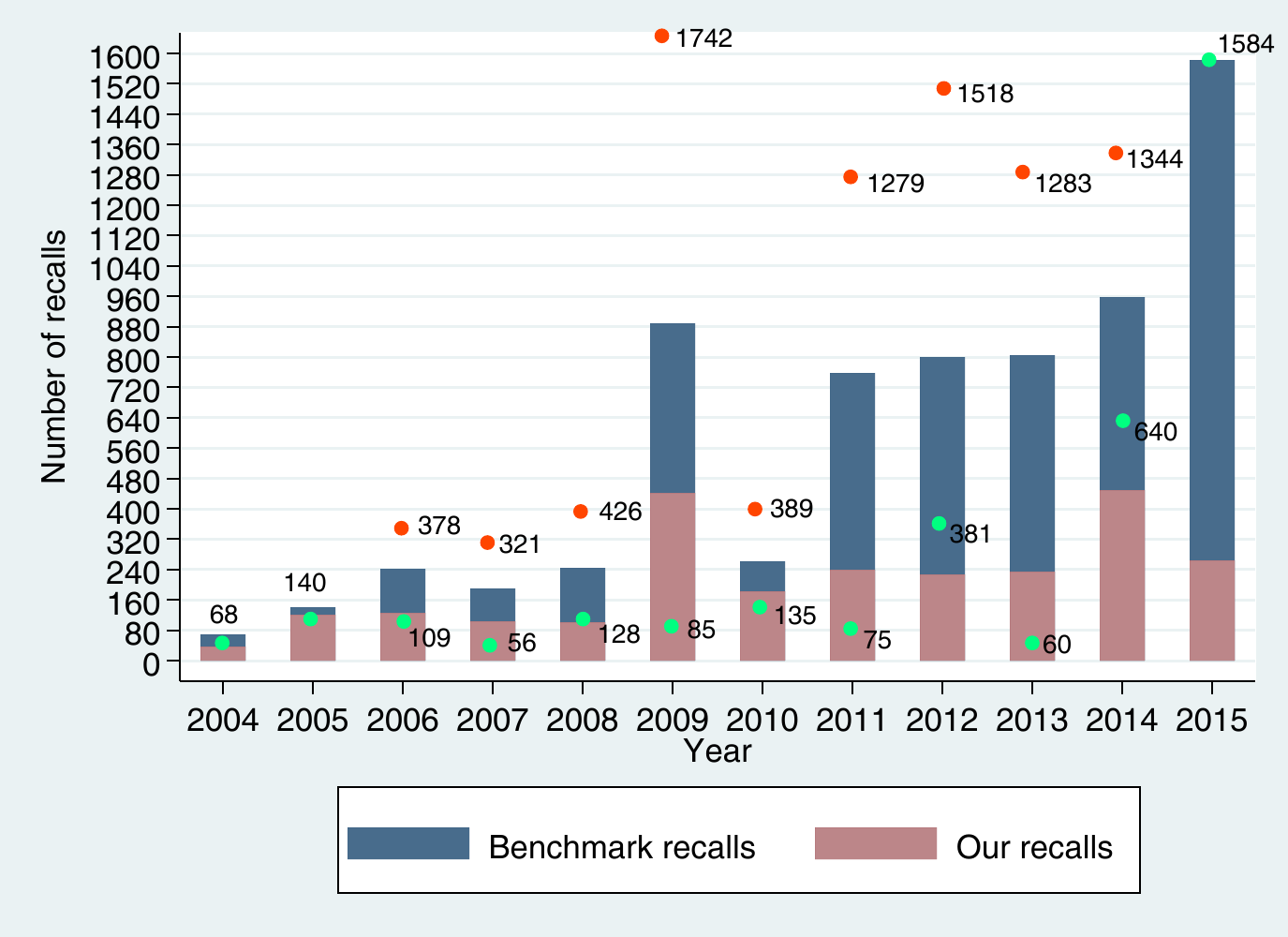}
  \caption{\footnotesize The number of our recalls against the number of recalls used as benchmark (average of sources). We include the minimum and the maximum number of recalls retrieved by the different sources. Mint colored points represent the minimum amount of recalls retrieved among all the sources at our disposal. Red points represent the maximum number of recalls among all the sources. A single mint point has been put whenever a single source was present for a year (2004, 2005, 2015).}
  \label{number_recalls_our_vs_b}
\end{figure}

\begin{figure}[H]
\centering
  \includegraphics[width=.8\linewidth]{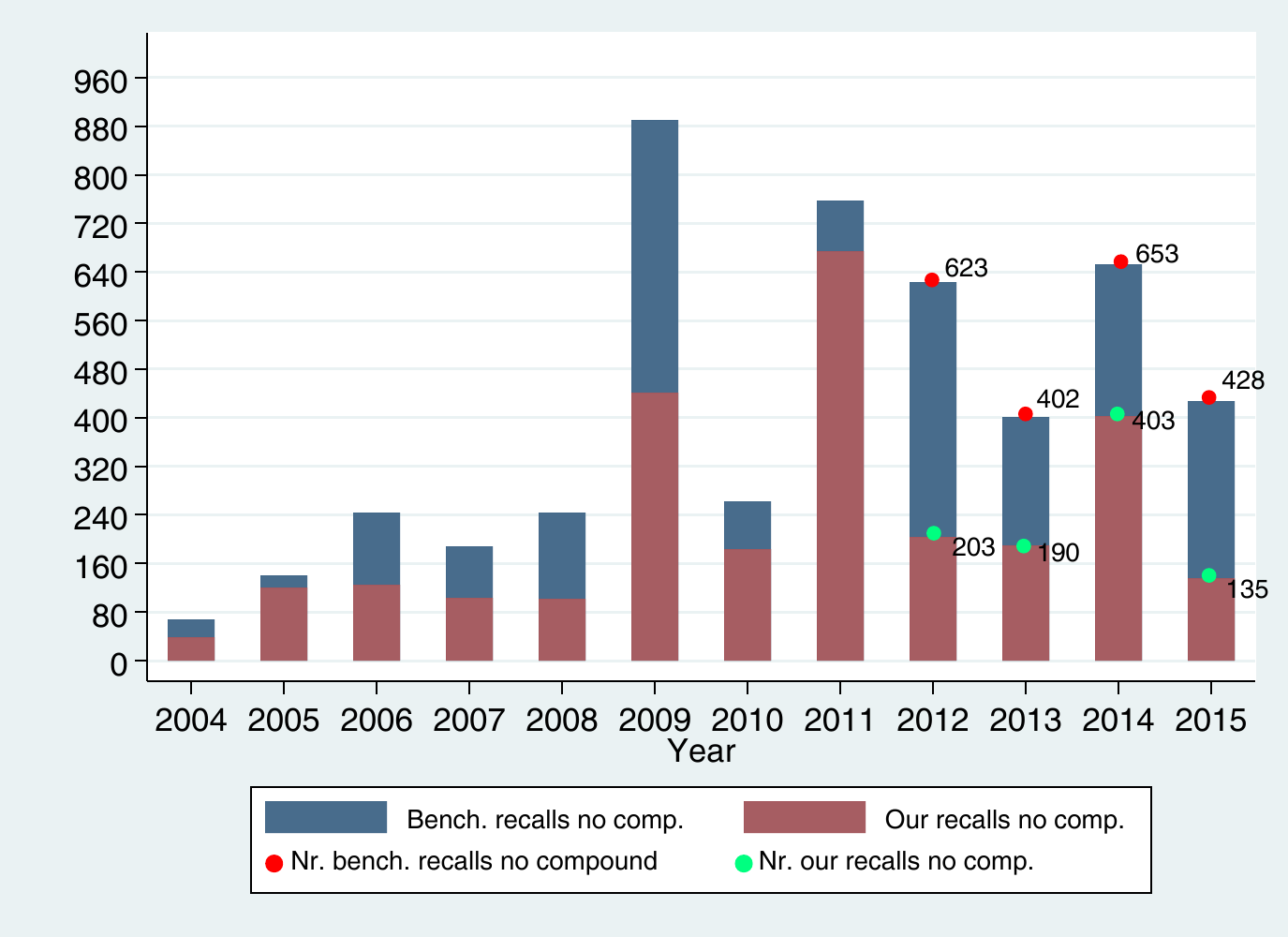}
  \caption{\footnotesize The number of our recalls against the number of recalls used as benchmaark without compounding recalls. We included the minimum and the maximum number of recalls retrieved by the different sources. Mint colored points represent the minimum amount of recalls retrieved among all the sources at our disposal. Red points represent the maximum number of recalls among all the sources.The situation i almost unchanged with respect to Fig. \ref{number_recalls_our_vs_b} until 2011. From 2011 on, the recalls collected in our dataset follow the benchmark if compounders' recalls are excluded more precisely.}
  \label{bench_no_compound}
\end{figure}

Fig. \ref{number_recalls_our_vs_b} underlines a disproportion in terms of the number of recalls starting from 2009, with respect to the benchmark. Such deficiency pertains to the counting methodology together with the structure of the database (see above).\\
Though the global trend is approximately reproduced, 2011, 2013, and 2015 represent problematic years. The dissimilarity of 2011 concerning the benchmark can be easily explained. Indeed, with the exclusion of Aidapak's recalls from the count of the collected recalls, the latter dropped. Furthermore, 2013 and 2015 have far fewer recalls than expected because more than 60\% of the recalls in 2013 and nearly 75\% of the recalls in 2015 were represented by compounding firms.\\
The recalls trend of the benchmark seems to be well reproduced. However, when recalls of pure compounders are excluded from the benchmark number and the sample of collected recalls. Fig. \ref{bench_no_compound} shows, indeed, an accordance in trends. Aidapak's recalls are here included. Indeed, Aidapak is a repackager and not a compounder. Besides, we want to show that 2011 does not constitute a problematic year once Aidapak's recalls are considered.

\begin{figure}[H]%
\centering
\includegraphics[width=16cm]{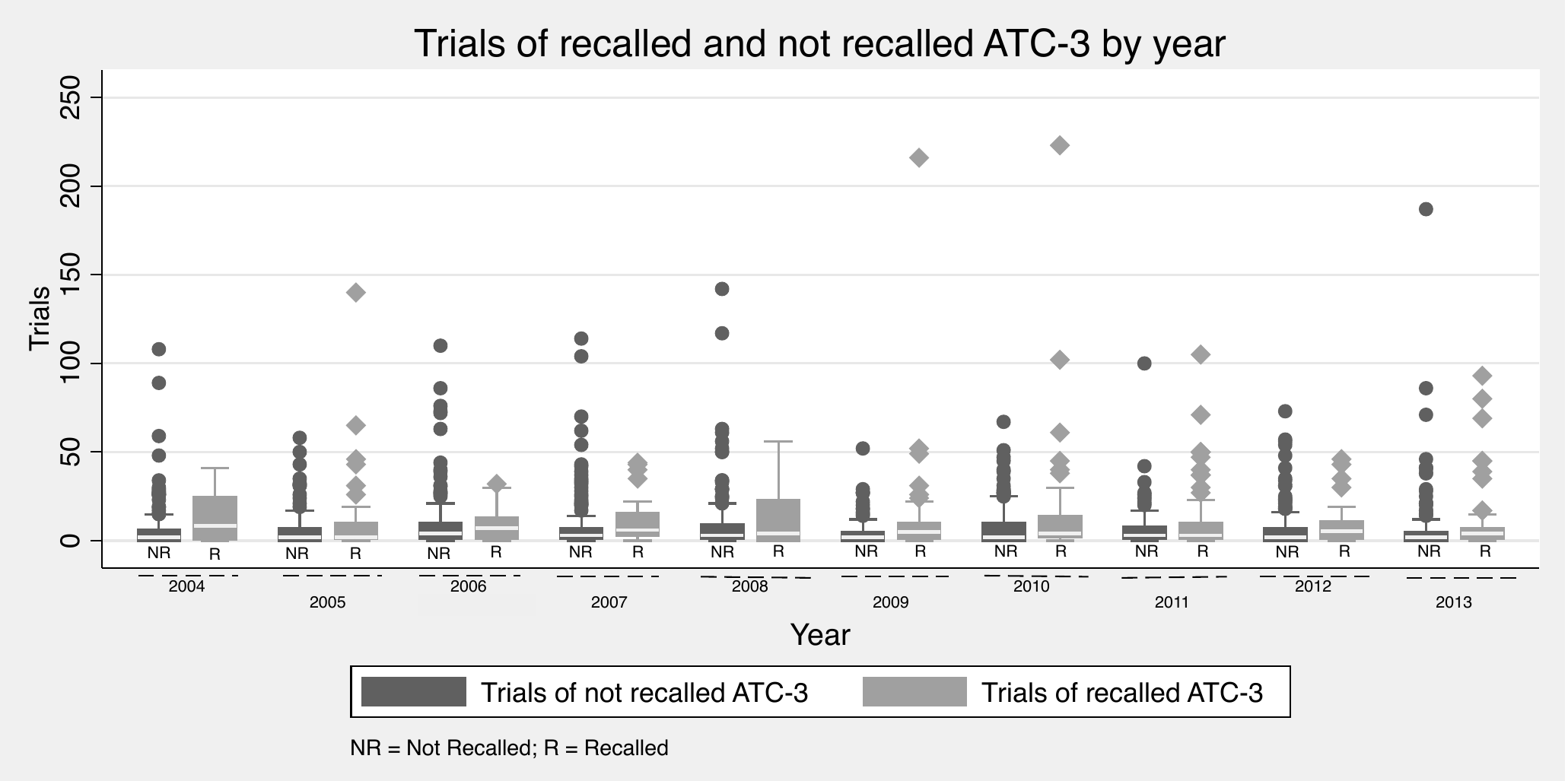}
\caption{The box plots show how, on average, recalls do not necessarily happen in more innovative markets. The average number of trials amounts to 123 for ATC-3 markets having undergone a recall and to 118 for ATC-3 markets not having undergone a recall.\\
The graph reports a yearly analysis of the average number of trials in recalled and not recalled ATC-3 groups. The average number of trials is similar in both the ATC-3 markets, having undergone at least a recall (R) and ATC-3 markets not having undergone any recall (NR).}
    \label{boxpl}%
\end{figure}

To conclude, as a further check of the exogeneity of recalls, we constructed a box plot displaying the average number of trials (and their dispersion) in both ATC markets having undergone a recall and not having undergone a recall by year. The latter exercise helps in understanding that major recalls do not necessarily intervene in more innovative markets. Indeed, Fig. \ref{boxpl} displays that the yearly number of trials of ATC markets undergoing major recalls almost coincides with the average number of trials in all other markets. 

\newpage
\section{Methodology} \label{methods}
The main theoretical framework is the same adopted in \cite{acemoglu2004market}. In particular, \cite{acemoglu2004market} model innovation, the dependent variable of the present model, as being proportional to market size. Refer to \cite{acemoglu2004market} for further details.\\
The measure of innovation is the number of clinical trials in all Phases for the ATC-3 category $i$. The measure of market size is the sum of products' sales for the $i^{th}$ market. When other potential determinants, time effects, and category effects are added to the analysis, the well-known estimation Poisson model is returned
\begin{equation}
\label{acemoglu}
E[N_{it}|\mu_i, \zeta_t, X_{it}, M_{it}] = \text{exp}(\beta_1 \cdot \text{log}M_{it}+\beta_2 \cdot X_{it} + \mu_i + \zeta_t) \quad \forall i = 1, \dots N, t = \, \dots T
\end{equation}
where $E$ is the expectations operator, $M_{it}$ represents endogenous market size, $X_{it}$ captures age (e.g., the average age of products in category $i$ weighted for products' size) , diversification and innovation patterns (e.g., the scientific production), $\mu_i$ are ATC fixed effects and $\zeta_t$ time fixed effects. 
The estimation of (\ref{acemoglu}) would lead to biased estimates for two reasons: first of all, the non-linearity in (\ref{acemoglu}) makes it impossible to estimate the fixed effects consistently; secondly, market size is endogenous.\\
In order to deal with both problems, a novel control function (CF) IV approach, described in \cite{lin2019testing} has been adopted. With respect to past literature, the present method allows to deal (i.e., testing and estimating) simultaneously with two potential sources of endogeneity: that due to correlation of covariates with time-constant, unobserved heterogeneity and that due to correlation of covariates with time-varying idiosyncratic errors. Furthermore, it can be easily extended to non-linear scenarios with fixed effects.\\
Specifically, denoting as $\kappa_{it}$ the idiosyncratic shock and $c_i$ the individual heterogeneity, the unobserved effects non-linear model allowing for both idiosyncratic endogeneity and heterogeneity endogeneity might look as follows: 
\begin{equation}
    E[N_{it}|M_{it}, z_{it}, c_i, \kappa_{it}] = c_i \text{exp}(x_{it}\beta_1+\kappa_{it})
\end{equation}
where $x_{it} =  (M_{it}, z_{it})$. $z_{it}$ would typically include a full set of time effects, and $M_{it}$ is the endogenous variable. All exogenous variables, which include the vector $z_{it}$ can be correlated with the heterogeneity (i.e., no random effects). There is also a set of excluded exogenous $R_{it2}$ serving as an instrument for the potentially endogenous variable. In the present work, $R_{it2}$ is represented by recalls. \cite{lin2019testing} noticed that, without the idiosyncratic endogeneity, an appealing estimator would be a fixed-effects Poisson estimator, which, viewed as a QMLE, would only require a strict exogeneity assumption with respect to the idiosyncratic shocks to ensure consistency. Such an assumption is exploited as a null hypothesis for testing idiosyncratic endogeneity against the alternative of full dependence of the error term of the specification of $M_{it}$ and $\kappa_{it}$. The alternative is composed by exploiting the reduced form equation for the endogenous variable 
\begin{equation}
M_{it} = z_{it}\Pi+c_{i2}+u_{it2} \quad \forall t = 1, \dots T
\end{equation}
where because the $z_{it}$ is strictly exogenous, it is tested the correlation between $\kappa_{it}$ and
functions of $u_{it2}$.  \cite{lin2019testing} developed a simple procedure allowing to test for idiosyncratic endogeneity and produce consistent estimates also in co-presence of non-linearity, fixed effects and both types of endogeneity. The algorithm follows the steps below:
\begin{enumerate}
    \item Estimate the reduce form for the endogenous through fixed effects and obtain the fixed effects residuals $\ddot{u}_{it2} = \ddot{M}_{it} - \ddot{z}_{it}\hat{\Pi} $
    \item Use fixed effects Poisson on the mean function $$E[N_{it}|M_{it}, z_{it}, c_i, \ddot{u}_{it2}] =c_i \text{exp}(x_{it}\beta_1+\ddot{u}_{it2}\rho)$$
    use robust Wald test of $H_0: \rho = 0$
\end{enumerate}

Step 2 allows estimating the fixed effects in the presence of non-linearity consistently. Yet, fixed effects Poisson enables eliminating ATC-level fixed effects performing a conditional ML consistent estimation. Refer to \cite{cameron2013regression} for further details.\\
A crucial characteristic of Poisson-FE models is that they require the dependent variable to be nonzero for at least one time period. The lower the proportion of zeroes in the dependent, the better the model works. The last condition has been fulfilled by dropping those ATC categories not meeting it, constituting approximately 10\% of the total ATC-3 in the sample. \\
We estimated several instances common to literature to check for either delayed effects of trials or the presence of a bias if market size were considered exogenous (or fixed effects omitted). \\~\\
Throughout, the problem of endogeneity in market size has been exposed as being intrinsic to market size. Hence instrumentation of the endogenous $M_{it}$ is needed. Market size is instrumented through normalized recalls. The normalization is on the number of products present in the market $i$ at time $t$. Calling $m$ the major recalls, normalized recalls are denoted as follows: $$\tilde{m} = \frac{m}{\# prod.}\cdot 100.$$ As aforementioned, normalization is necessary in order to avoid another source of endogeneity. Indeed, ATC markets having more products are more likely to undergo a recall by definition. Omitting such control would partly invalidate the estimates. The belief is that markets undergoing major recalls experiment with a sudden negative shock in sales.
The relevance of the instrument is tested in Section \ref{resu}.\\
The instrument is not directly related to the dependent variable. The central argumentation that might directly connect normalized recalls to trials is that the lack left by recalls is filled with innovations. Hence sectors more prone to undergo a recall should also be the most innovative ones. In literature, there seems to be contrasting evidence about the topic. Though the argumentation would imply a positive impact of recalls on innovation, the recent events seem to contradict such findings. Indeed, albeit an increasing number of recalls from 2004 to 2015 (see, e.g., Fig.\ref{number_recalls_our_vs_b}), the innovation crisis of the pharmaceutical industry is a widely known and recognized phenomenon in literature (see e.g.\cite{pammolli2011productivity}, \cite{price2014making} among others). It might be argued that the contrasting effects leading to the drop of innovation have overtaken the positive effect of recalls, thus favoring the decreasing pharmaceutical innovation trend. The positive effect of recalls on innovation could be, therefore, still present but hidden. Empirical research has conducted few analyses to explore the relationship between innovation and recalls or withdrawal in general. Fortunately enough, most severe recalls have considerable media coverage, which allowed researchers to collect data on market reaction to such bad events (see, e.g.,  \cite{perez2012product}). Authors working on such a stream of literature conclude that the impact of recalls and withdrawals on market innovation has a high variability: some recalls have considerable effects while others have none at all. There seems not to be a systematic way to identify the recalls whose announcement impacted innovation among major recalls. Market reactions depend on not controllable criteria, such as the period during which the recall took place and eventual delays in the FDA's communication of the recall.
Generally, however, the market does not systematically overreact to such shocks, invalidating any dependence between recalls and innovation.\\
To summarize, direct connection sources between innovation and recalls are mainly due to fixed and time effects. FDA delays cannot be easily controlled. The FDA developed precise guidance and protocols for recall communication and announcement for the period considered in the present work. Hence, delays constitute a minor issue because FDA regulates them. For the sake of completeness, thanks to the FOIA agreement signed, openFDA, and FDA Enforcement report, it has been possible to verify the happening of delays. The mentioned sources allowed us to access the time gaps between recall initiation, recall classification, and recall termination. The communication of the recall is part of the initiation process. Above all, in case of severe recalls, it must be prompt. The average time between the initiation and the termination for Class I and Class II recalls has been around 23 months. A delay in communication might happen in the first initiation phase. The average time that the initiation phase took for any Class I and Class II recall was four months approximately. For our sample of major recalls, the initiation phase's average time has been approximately 2 to 3 months, in line with prompt communication criteria. This evidence enforces the limited impact of delays on the analysis.  \\
Dropping out unobserved heterogeneity and including time dummies in the primary specification control for possible direct connections between recalls and innovation. Thus, the mentioned operations ensure only an indirect effect of recalls through sales. \\
Further arguments in favor of the indirect effect of recalls on innovation follow.\\
In particular, the recalls taken into account are severe recalls of marketed products. The time gap between trial phases and the marketing of a drug usually takes between 8 to 14 years. Such a significant time gap is relevant to guess and understand competitors' possible reactions to a drug recall in the same sector where a firm is operating. We believe that a competitor that underwent a recall in the sector in which both firms operate does not increase or decrease the risk of innovation in the short run. Indeed, marketed products undergo major recalls long after that they are commercialized. \\
Besides, the lack of sales left on the market by recalling the drug requires an extended period to recover fully. Hence, there is no need to invest in clinical trials to take advantage of such a shortage in the short run.  As a further check f the latter conjecture, we build up a time-to-event analysis in Fig.\ref{km_sur} of Appendix. Fig.\ref{km_sur} takes into account all types of recall and clearly shows how a recalled product has a truncated life compared to drugs having a normal life cycle.\\ 
In particular, Fig.\ref{km_sur} displays how the survival rate of drugs that did not undergo a recall is persistently higher than the survival rate of drugs having undergone a recall. Hence, having undergone a recall decreases the "probability of surviving of a drug." Under normal conditions, drugs have a probability greater than 0 to survive more than ten years. However, if a drug underwent a recall, this probability drastically reduces to almost 0. Notice that the probability that a recalled drug survives two years is still consistent. The median survival time is five years.\\ 
Thus, the drop in sales after a recall is likely to remain unfilled for years. Indeed, had firms found innovative replacements for recalled drug $d$, which allowed them to recover the shortages left by the recall of $d$, there would not be any reason to keep selling drug $d$ for years. Recalled products, therefore, leave a long-term lack in terms of sales within the ATC-3 market to which they belong.\\
A further argument against the coverage of lacks left by recalls through innovative products is that such shortages might be filled by drugs already present in the market, whose trials started before, soon after, or at the same time as the trials leading to the recalled drug. This eventuality is reasonable since, as mentioned, suspended or terminated studies are excluded from the sample, meaning that remaining clinical trials sponsored by concurrent firms are likely to arrive on the market with products belonging to the same therapeutic class. Competition of wholesalers within an ATC might reveal in early stages once it is evident that a firm will develop an innovative cure. The development of alternative drugs is encouraged from the early trial phases when there are still chances to arrive first on the market. Medicines substituting recalled drugs in the same ATC might be developed soon after the recalled medicine in a "first to arrive" competition rather than a "fill the gaps of recalls" logic. The latter may also be because the demand for patented medicines of the type of the recalled drug was likely more consistent when the trial for the recalled drug started. In the eventuality that demand propagates at the recalls' time, either already existing generics or new ones (trials of generics is indeed less time consuming since they only need to ensure bio-comparability) might intervene and fill the gap.\\
It is worth noticing, in any case, that the potential positive relationship of recalls and innovation exploiting the market lacks passes indirectly through market size. Indeed, the emergence of new trials within a market after a recall depends on the demand that the product in question generated in the market. If a recalled product had no underlying demand, it is reasonable to expect no company to begin a costly trial only to fill the lack left by the recalled product. Therefore, the response of innovation seems to depend not directly on the recall but the underlying magnitude of the recalled product's demand, i.e., on market size.\\
Finally, another possible critique undermining the instrument's validity is that recall of product $i$ might have provoked the recall of trials concerning similar products. This domino effect hangs on the causes of the recall. Indeed, if the recall concerns only the specific product being withdrawn from the market, implications on other companies' products are unlikely. For instance, it is possible that after the recall of the COX-2 inhibitor, Vioxx, due to cardiovascular side effects, all firms having ongoing trials on the same target did suspend or withdraw the trials relating to COX-2 inhibitors. To the best of our knowledge, no effort has been made to explore this possibility in the drugs market. The only work approaching the critique is \cite{ball2018recalls}. The author, however, focuses on the medical devices industry, which has different legislation for recalls than the drugs' market. Indeed, a device's recall is a common practice made ordinarily by firms to repair or update a device, which is usually promptly placed back to the market. The way we managed the circumstance is threefold. First, we considered only active trials, thus excluding suspended and withdrawn trials, including those suspended due to other drugs' recall. In a second instance, we removed trials of companies undergoing a recall as well. Ultimately, as far as it has been possible to link the reason of the severe recalls \footnote{above all, in case of adverse events caused by an active principle adopted in the drug to the scope of a trial} we dropped trials adopting a similar active principle. The latter instance happened in a few cases since eliminating suspended and withdrawn trials constitutes already a robust control.

\newpage
\section{Results} \label{resu}
The results section is divided into two main subsections. Namely, the impact of recalls on the endogenous market size is first analyzed as measured by total sales of ATC $i$. The aim is to provide convincing arguments in favor of the relevance of the adopted instrument. \\
Successively, are presented the results of the impact of the instrumented market size on innovation. 
\subsection{The impact of recalls on sales}
\subsubsection{Summary statistics}\label{sec:descriptive}
This Section reports summary statistics for the sample. Tab.\ref{sumstats} contains average values and standard deviations (below) of relevant variables for the full sample and two separate sub-samples for observations associated or not to recalls. The Table includes such information at the ATC-3 level and refers to major recalls.Tab.\ref{sumstats} embraces all the relevant controls employed for constructing Tab.\ref{fiststage}.

\begin{table}[H]
\centering
  \caption{\footnotesize Summary statistics at ATC-3 level for the full sample, the subset of ATC-3 having undergone a recall 
  in the period considered, and the subset not having undergone a recall. Database at ATC-3 level is balanced.}
\makebox[\textwidth][c]{
\scalebox{0.6}{
    \begin{tabular}{cc|ccc|c}
\toprule
     &  & \multicolumn{3}{|c|}{\textbf{ATC-3}}\\
\cmidrule(lr){3-5}
 \textbf{Variable} & & Full Sample  & Subs. recalls & Subs. no recalls & \textbf{Description}\\ \midrule
\multirow{4}{*}{Sales (log)} & Overall mean & 19.405 & 20.794 & 19.054  & \multirow{4}{*}{Log of sales at ATC-3 level.}\\
& Overall Std. Dev. & 2.233 & 1.475 & 2.256  \\
& Between Std. Dev.  & 2.152 & 1.441 & 2.163  \\
& Within Std. Dev. & .614 & .378 & .661  \\
\midrule
\multirow{4}{*}{Outflow rate ($\frac{K_{t+1}}{P_{-1}}$)} & Overall mean & .086 & .056 & .093  & \multirow{4}{*}{\thead{It is defined as the number of lost products in  \\ an ATC-3 ($K_{t+1}$ in regressions) over the total \\ number of products in $t-1$ ($P_{-1}$ in regressions). }}\\
& Overall Std. Dev. & .257 & .064 & .285  \\
& Between Std. Dev.  & .109 & .036 & .120  \\
& Within Std. Dev. & .233 & .053 & .259  \\
\midrule
\multirow{4}{*}{\thead{Avg. age of firms \\ within ATC}} & Overall mean  & 35.907 & 33.275 & 36.573 & \multirow{4}{*}{\thead{It is the average age of the firms competing \\ within an ATC-3. The foundation year of \\ the firms was present in the data.}} \\
& Overall Std. Dev. & 7.631 & 5.420 & 7.960  \\
& Between Std. Dev. & 6.767 & 4.452 & 7.093  \\
& Within Std. Dev.  & 3.556 & 3.160 & 3.650  \\
\midrule
\multirow{4}{*}{\thead{Herfindahl–\\Hirschman Index \\ (hhi)}} & Overall mean  & .431 & .268 & .434 & \multirow{4}{*}{\thead{The hhi measures the competition within a market. \\It can range from 0 to 1.0, moving from a huge number \\ of very small firms to a single monopolistic producer.}} \\
& Overall Std. Dev. & .260 & .159 & .262  \\
& Between Std. Dev. & .236 & .166 & .240  \\
& Within Std. Dev. & .110 & .020 & .115  \\
\midrule
\multirow{4}{*}{\thead{Share generics \\ by ATC}} & Overall mean  & .746 & .725 & .752 & \multirow{4}{*}{\thead{It represents the percentage of generic products, \\ among all products sold in an ATC-3 market}} \\
& Overall Std. Dev.  & .255 & .214 & .264  \\
& Between Std. Dev. & .238 & .210 & .245  \\
& Within Std. Dev. & .092 & .052 & .099  \\
\midrule
\multirow{4}{*}{\thead{Avg. age prod. \\ by ATC}} & Overall mean  & 13.159 & 12.043 & 13.441  & \multirow{4}{*}{\thead{It represents the average age of product within \\ an ATC-3. The age of a product is based on \\ the foundation year of the firm that produced it.}}\\
& Overall Std. Dev. & 5.333 & 3.763 & 5.627  \\
& Between Std. Dev. & 4.909 & 3.568 & 5.164  \\
& Within Std. Dev. & 2.109 & 1.304 & 2.268  \\
\midrule
\multirow{4}{*}{\thead{Scientific knowledge \\ within ATC}} & Overall mean &  6.327 & 6.787 & 6.211 & \multirow{4}{*}{\thead{The number of papers and scientific publications for \\ an ATC-3 present in PubMed and other sources.}} \\
& Overall Std. Dev.  & 1.718 & 1.623 & 1.724  \\
& Between Std. Dev. & 1.705 & 1.627 & 1.709  \\
& Within Std. Dev. & .242 & .177 & .256  \\
\midrule
\multirow{4}{*}{\thead{Number of firms \\ within ATC}} & Overall mean & 21.054 & 32.802 & 18.082 & \multirow{4}{*}{\thead{Number of firms trading within an ATC-3}} \\
& Overall Std. Dev. & 20.954 & 23.442 & 19.175  \\
& Between Std. Dev. & 20.604 & 23.069 & 18.876  \\
& Within Std. Dev. & 4.050 & 5.367 & 3.645  \\
\bottomrule
\end{tabular}}%
}
\label{sumstats}%
\end{table}%

Tab \ref{sumstats} displays the overall, between, and within standard deviation for the main controls included sales. The statistics are provided for the total sample, the subset of ATC-3 having undergone at least a recall, and the sub-sample of ATC-3 without recalls.  The panel of sales in Tab. \ref{sumstats}, displays how, typically, the recalls are found in larger markets than the average. For this reason, recalls have been normalized by the number of products in the ATC market to avoid possible problems of reverse causality with the market size. The normalized recalls have been denoted as $\tilde{recalls}$ in the following paragraphs. \\
Moreover, as expected, more competitive markets are more prone to recalls, as displayed by the Herfindahl–Hirschman Index (hhi). There is evidence of differences in terms of competition between ATC-3 groups.
\cite{nutarelli} provides further insights about which type of firms and products generally undergo a recall. Specifically, \cite{nutarelli} evidences a general tendency of recalls to be located in big established firms and to regard relatively older products than the average age. \\
On the contrary, with respect to firm and product levels,  recalls are located in more dynamic ATCs, where recalled drugs were pioneering in the past. Fixed effects technique accounts for time-invariant characteristics of ATCs.\\
The recalls intervene in firms with a high share of generics (see \cite{nutarelli}). This finding might result from a less stringent policy for generic drugs' approvals compared to branded ones. Growing concern for generic safety is, in fact, a well-known problem in literature (see, e.g., \cite{gallelli2013safety}).\\ 
Besides, in ATC markets, the outflow rate presents a within variance higher than the between variance. The latter means that there is no difference between ATC-3 groups concerning the outflow rate.
As opposed to the firm level, this inversion is expected. Indeed, while strategic policies of product placement might occur in firms, this is not the case for ATC aggregation, where market laws apply. Thus, on average, even two utterly different ATC markets would display similar outflow rates following only a demand-supply logic.\\
Two other variables seem to be related to recalls at the ATC-3 level, i.e., scientific knowledge within an ATC and the number of firms trading within an ATC.  Specifically, the recalls happen in ATC markets where, on average, trade more firms and scientific knowledge is more advanced than other markets.\\
To summarize, the recalls regard relatively old drugs produced in big established firms. The major recalls occur in relatively dynamic markets whereby, on average, many younger firms operate, trading relatively young products. A possible reason the markets having the described characteristics undergo more easily recalls is that they are precisely the markets monitored by the legislator with special attention.

\subsubsection{Analysis of the determinants of drug recalls}\label{sec:selection}
This section reports the first-stage results. A Fixed-Effects estimation method is employed. Tab.\ref{fiststage} shows the estimates of the first stage at the ATC-3 level. As detailed below, a further level ATC-Firm has been added to test for compensations within ATCs inside firms. For consistency with the best model, the sample was truncated in 2013 also for the first stage. Outcomes with a not-truncated sample display very similar results (see \cite{nutarelli}). The F-statistic amounts to 14.32. The standard errors included in the Tables of the present Section and the following ones are all robust and clustered at the ATC-3 level of aggregation.

\begin{table}[htbp]\centering
\def\sym#1{\ifmmode^{#1}\else\(^{#1}\)\fi}
\captionsetup{justification=centering}
\caption{\footnotesize First stage results at different levels. \\ ATC-3 aggregation represents the main specification. \label{fiststage}}
\scalebox{0.6}{ \vspace*{-0.01mm}\begin{tabular}{l*{2}{D{.}{.}{-1}}}
\toprule & \multicolumn{1}{c}{\textbf{(ATC-Firm Aggregation)}} & \multicolumn{1}{c}{\textbf{(ATC-3 Aggregation)}} \\
                         &\multicolumn{1}{c}{Log sales}&\multicolumn{1}{c}{Log sales}\\
\midrule
$\tilde{recalls}$         &     -0.0053         &     -0.0283\sym{***}\\
                          &    (0.0033)         &    (0.0056)         \\
$\tilde{recalls}_{t-1}$   &     -0.0267\sym{**} &     -0.0267\sym{***}\\
                          &    (0.0083)         &    (0.0070)         \\
$\frac{K_{t+1}}{P_{-1}}$ &                     &      0.1932\sym{**} \\
                          &                     &    (0.0628)         \\
\textit{average age firm}
&                     &      0.1576         \\
                         
                          &                     &    (0.0921)         \\
\textit{average age firm}$^{2}$ &                     &     -0.0020         \\
                      &                     &    (0.0013)         \\
hhi     &                     &      1.2405\sym{***}\\
                 &                     &    (0.2590)         \\
\textit{share generics in ATC}
&                     &     -0.1895         \\
                         
                          &                     &    (0.3373)         \\
papers           &                     &     -0.0260         \\
                  &                     &    (0.0507)         \\
\textit{\# firms}        
&                     &      0.0077         \\
                        
                         &                     &    (0.0071)         \\
\multicolumn{1}{l}{Year Dummies} & \multicolumn{1}{c}{Yes} & \multicolumn{1}{c}{Yes} \\                                            \midrule  \multicolumn{1}{l}{Obs.}  & \multicolumn{1}{c}{48915} & \multicolumn{1}{c}{1664} \\                                                                                        \multicolumn{1}{l}{Groups} & \multicolumn{1}{c}{8634} & \multicolumn{1}{c}{208} \\
\bottomrule
\multicolumn{2}{l}{\footnotesize Standard errors in parentheses}\\
\multicolumn{2}{l}{\footnotesize \sym{*} \(p<0.05\), \sym{**} \(p<0.01\), \sym{***} \(p<0.001\)}
\end{tabular}}\begin{spacing}{0.1}\vspace{0.3cm} \footnotesize  {\fontsize{50}{60}\selectfont}{\fontsize{50}{60}\selectfont}{\fontsize{5}{6}\selectfont}{\Huge }{\begin{minipage}{11.6cm}\tiny Huber-White robust and clustered at ATC-3 level standard errors are in parentheses. First-stage results are shown in this Table. (1) fits an F.E. model at ATC-Firm level, i.e., ATC lines of productions within firms. This level is introduced to check the possibility of compensations between sales of products belonging to the same ATC (excluded due to the significance of the coefficient of recalls witnessing a drop after a recall) (2) fits an F.E. model at the ATC-3 level. $\tilde{recalls}$ represent recalls normalized. At the ATC level, recalls are respectively normalized for the number of products within an ATC. \end{minipage}} \\ {\fontsize{50}{60}\selectfont}{\fontsize{5}{6}\selectfont}  \end{spacing}
\end{table}

\vspace{0.2 cm}
We found a significant and negative impact of recalls on the logarithm of sales at the market level. In \cite{nutarelli} it is shown that sales of firms undergoing a recall are unaffected.
At the same time, the prouction lines of medicines belonging to the same ATC-3 encounter a drop in sales due to recalls (ATC-Firm Aggregation in Tab.\ref{fiststage}). This evidence excludes the possibility of compensations between sales of products belonging to the same ATC inside a firm.  Therefore, the negative effect of recalls at the market level is enforced, whose lacks are not filled by the same firms with other medicines of the same ATC-3.  \\
The second column of Tab.\ref{fiststage} represents the first stage of the principle analysis. As illustrated, the effect of recalls at the ATC-3 level is powerful and significant for current recalls and delayed ones. After having performed a sufficient amount of bootstrap repetitions, we found that the t-statistic is invariant to whether we use recalls or lag recalls to obtain it \footnote{t-stat is obtained after 30000 repetitions and amounts to  2.438}. This finding corresponds to a Sargan-Hansen test for over-identification in our contest, implying the absence of over-identifying restrictions (\cite{lin2019testing}). \\
We believe that the key reason for the strength of the result relies on the level of aggregation. While firms with high-quality managements and inclined to risk can promptly make up for severe recalls, the latter take ATC-3 markets unaware. Competitors could not anticipate severe recalls against firms producing in the same ATC as theirs, which can be detected only at the market level. \\
The absence of compensations at the market level has been further tested. In particular, we analyzed the effect of recalls on aggregated sales once the firms' sales having undergone a recall are removed from the sample. The drop in sales seems to disappear once firms having undergone a recall are excluded (see Fig. \ref{sales_nowithd} in Appendix).  \\
The fall of sales observed at the ATC-3  level becomes evident not only from the estimates in Tab.\ref{fiststage} but also from the study of abnormal values in Section \ref{abngrowth}\\ 
Finally, it might be argued that since recalled products are the most innovative ones, no direct substitute is present in the same market. However, having the generic name of products (both recalled and not recalled) and the active principle of medicines, it has been possible to detect an average of 10 products within the market exploiting the same active principle as the recalled products. Hence, it has also validated the hypothesis that the lacks left by recalls are probably filled with products already present on the market and that recalled products are not necessarily the most innovative ones having no substitutes.\\

\subsubsection{Analysis of Abnormal Values}\label{abngrowth}
This Section reports estimates of the influence of drug recall on sales. The effect of recalls is defined by taking a reference value of the given economic indicator as it would be observed under ``normal'' dynamics of economic conditions; this is called the ``potential'' value. We hence define the Abnormal Value (AV) of the indicator $y$ associated with the unit $i$ in time $t$ as the observed and potential value difference.
\cite{ThirumalaiSinha2011}:
\begin{equation}\label{eq:abnv}
AV_{it}=y_{it}-E\left(y_{it}\right),
\end{equation}
The potential value $E(y_{it})$ is estimated by running a Fixed-Effects regression on the following model:
\begin{equation}\label{eq:exp}
y_{it}=\alpha + \beta y_{st} + \gamma X_{it} + \mu_{i} + \lambda_{t} + u_{it},
\end{equation}%
where $y_{st}$ is the aggregated value of $y$ in year $t$  at the sector level. Usual control variables ($X$) and year dummies are included as regressors. After obtaining estimates of $AV_{it}$ for all $i$ and $t$, referred to as $\widehat{AV}_{it}$, the time dimension is re-scaled. Specifically, the time dimension is centered on the year when the recall is issued for all units experiencing a recall in the time frame considered. Only these units are kept in the sample. The market-level Abnormal Value $\overline{AV}_{t}$ associated to recalls is then computed as the simple average of $\widehat{AV}_{it}$ for any $t\in \{-(T-1),...,(T-1)\}$, as follows:
\begin{equation}\label{eq:avagg}
\overline{AV_{t}}=\sum_{i=1}^{N_{t}} \widehat{AV}_{it},
\end{equation}
where $N_{t}$ is the number of units with available data in $t$ among those experiencing one recall. Confidence intervals for $\overline{AV}_{t}$ are constructed calculating the variance of $\widehat{AV}_{it}$ as follows:
\begin{equation}\label{eq:avaggvar}
Var\left(\overline{AV}_{t}\right)=\frac{\sum_{i=1}^{N_{t}} Var\left(\widehat{AV}_{it}\right)}{N_{t}^{2}},
\end{equation}
where $Var(\widehat{AV}_{it})$ is the variance of the forecast error derived from estimation of Equation \ref{eq:exp}. The focus of the analysis is on the growth rate of sales volumes. The exercise is replicated for three classifications of recalls (standard recall definition, major recalls, type of recall) and three levels of analysis: product, firm, and sector level. The main text reports only the analysis at the ATC-3 level as it is the level at which the first and second stages are conducted. Abnormal values at the firm and product level can be found in Appendix.\\ Note that in the model for the sector level, $y_{st}$ is replaced with $y_{mt}$ in Equation~\ref{eq:exp}, that is the value at the whole market level. \\

Fig. 5 reports estimates of the effects of recalls on the AV of sales growth.

\vspace{0.5cm}
\begin{figure}[H]
  \centering
   \begin{adjustbox}{minipage=\linewidth,scale=0.8}
\hfil\hfil\includegraphics[width=7cm]{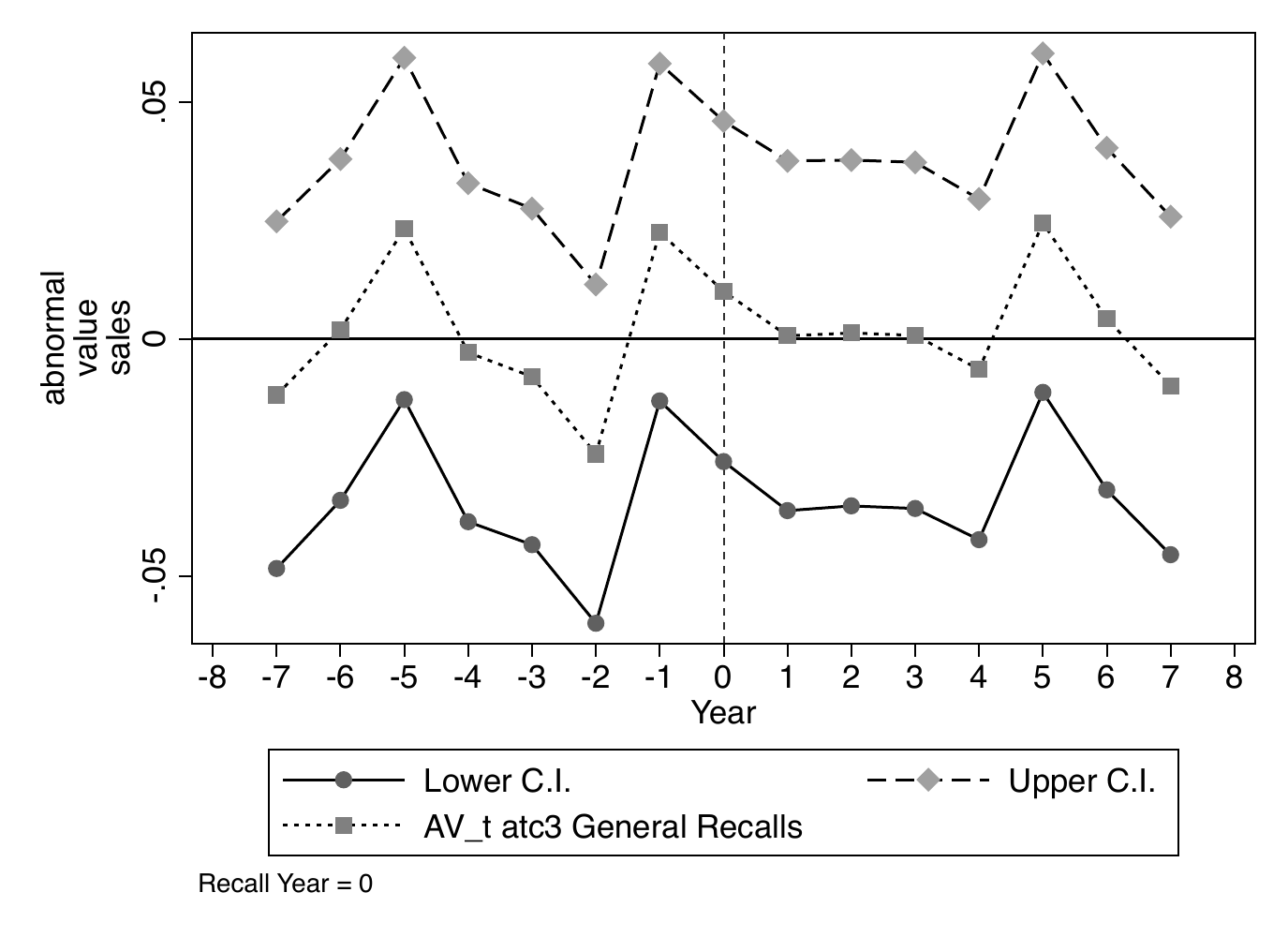}\hfil\hfil
\includegraphics[width=7cm]{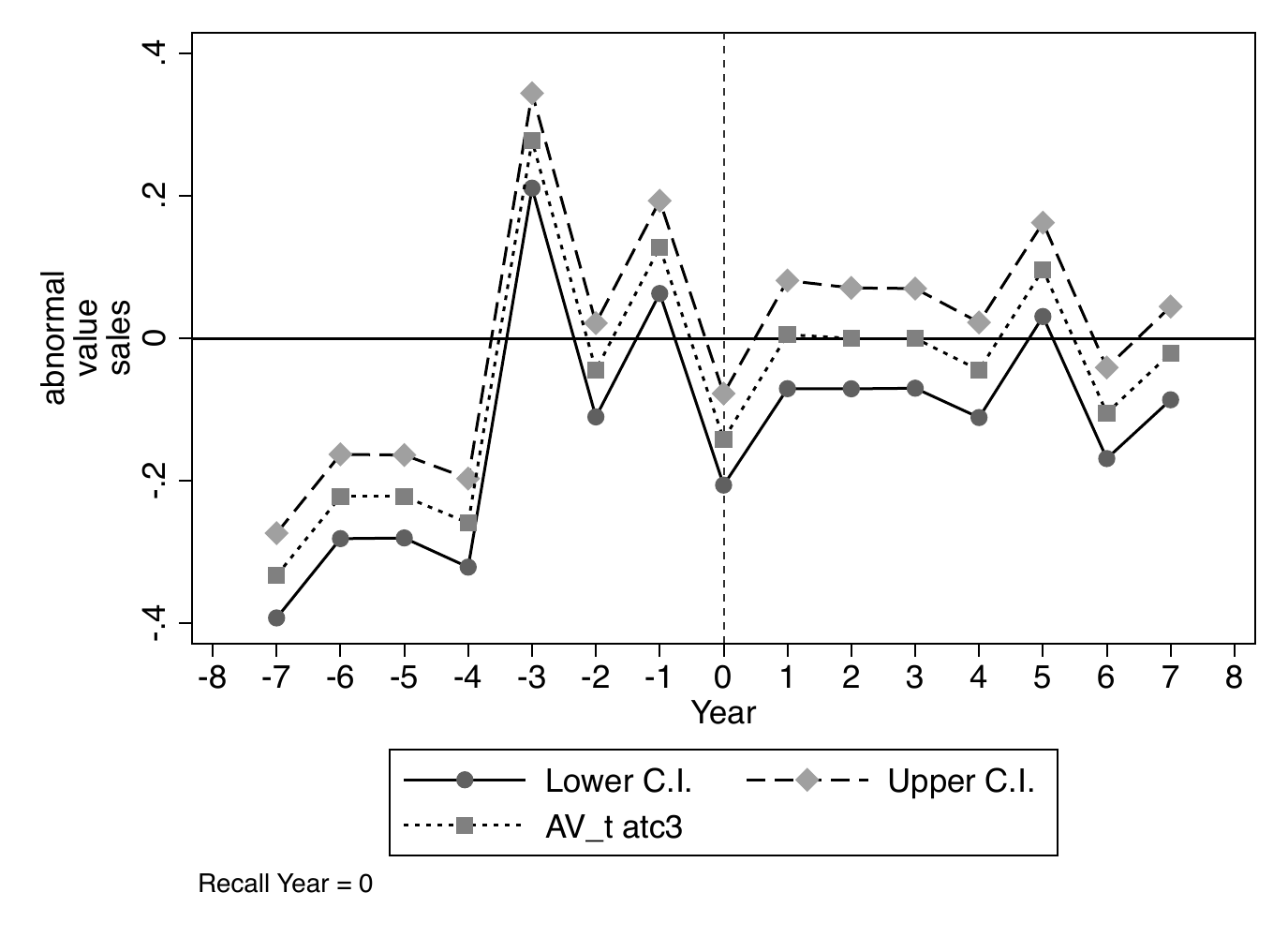}\newline
  \null\hfil\hfil\makebox[5cm]{ATC-3: general recalls sales}
  \null\hfil\hfil\makebox[5cm]{ATC-3: Maj. recalls sales}
  \vfil
  \hfil\hfil\includegraphics[width=7cm]{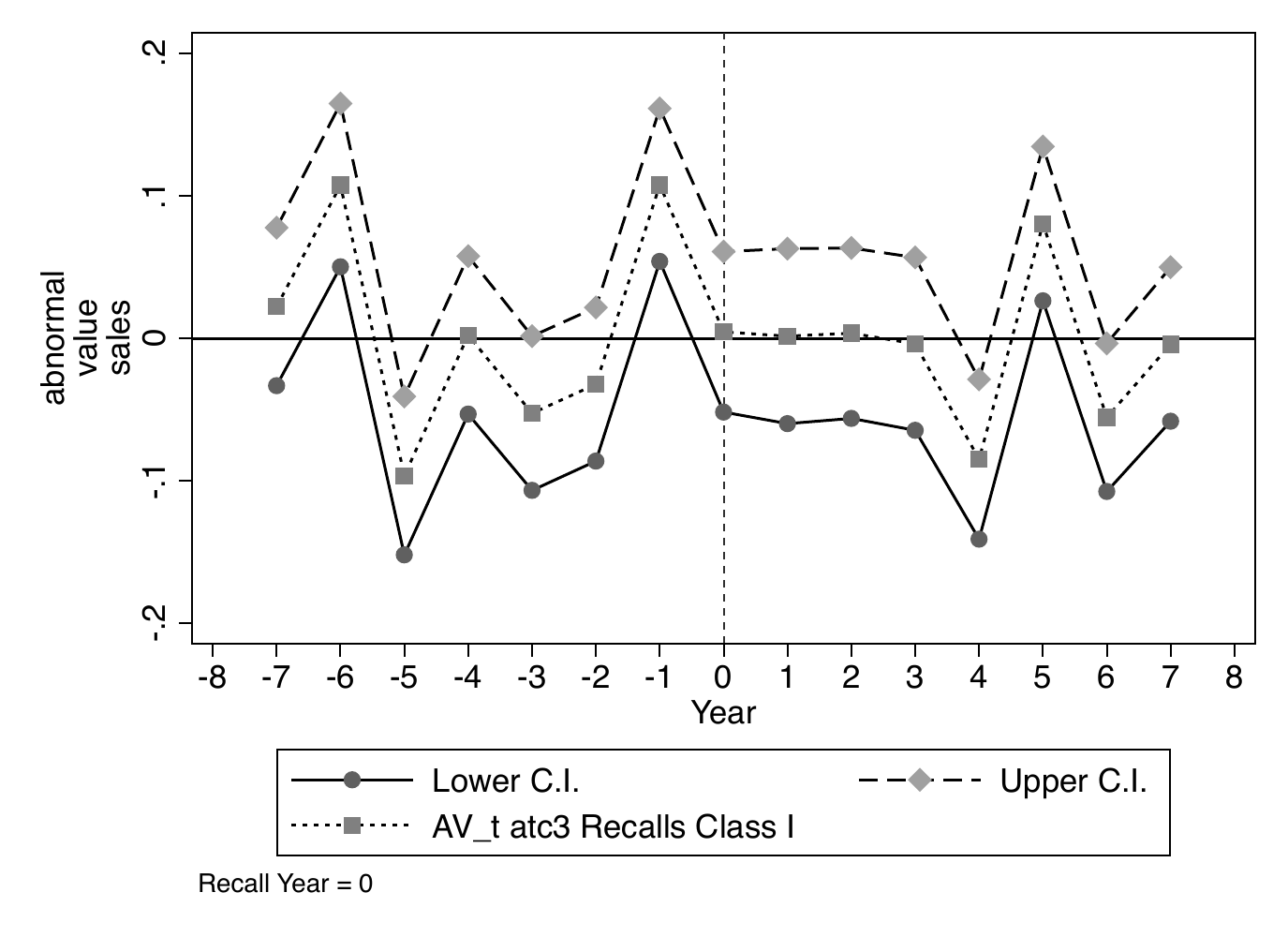}\hfil\hfil
    \includegraphics[width=7cm]{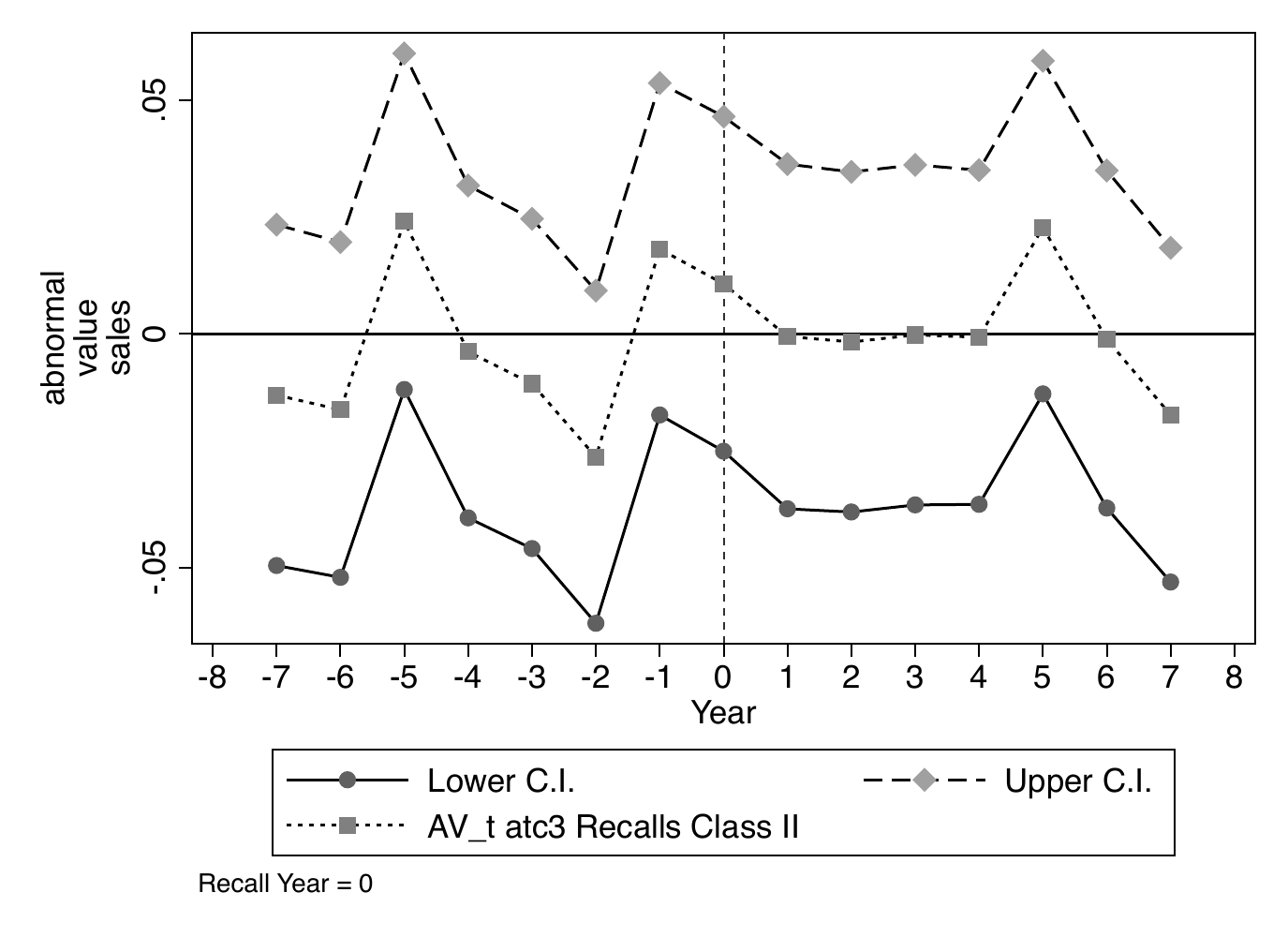}\newline
  \null\hfil\hfil\makebox[5cm]{ATC-3: Class I recalls sales}
    \hfil\hfil\makebox[5cm]{ATC-3: Class II recalls sales}
  \caption{Abnormal values at ATC-3 level of aggregation. Years are normalized. Year 0 represents the year of recall. The four scenarios include the path of sales before and after the recall year, using four different definitions of recalls: major recalls, Class I recalls, general recalls and Class II recalls. As it is evident from the diagrams, sales drop at recall year for every type of recall. Major recalls present a more pronounced drop. Moreover, using major recalls the lowest error bound is reached.}
  \label{atc3_sales}
  \end{adjustbox}
\end{figure}


Fig.\ref{atc3_sales} exhibits abnormal values for ATC-3 level. Confidence intervals are constructed at the 95\% level. As it is evident from Fig.\ref{atc3_sales}, after the initial drop at the year of recall, sales soon recover one or two years after year 0 (see major recalls). The latter observation classifies the instrument employed in our work as a short-run effect. This distinguishes the effect of our instrument from the long-run effect that demographic shocks produce in the work of \cite{acemoglu2004market}.\\
The analysis of abnormal values confirms what was found in previous paragraphs. Especially a considerable impact of recalls on sales in the year of the recall. The error bound is lower for the ATC-3 level with major recalls, thus enforcing the expectation of a drop in sales at the recall time.


\subsection{Relation between innovation and market size}
 In this section we report the results concerning the relationship between market size, $M_{it}$ and innovation, $N_{it}$. Since the data at our disposal are already converted into dollars of 2015 using Consumer Price Index (CPI), market size is measured directly as the sum of sales over ATC market $i$ at time $t$. Innovation is measured with the number of activated trials in ATC $i$ at time $t$. The time window ranges from 2004 to 2013. The samples' last two years (2014, 2015) have been cut away since very few trials have been conducted in such a period. Including 2014 and 2015 may have led to biases in the procedure, which exploits Poisson estimates. Indeed, the latter method does not tolerate a value of 0 for the dependent in most observations. \\
The panel is strongly balanced as required by the procedure. Each year has data for 208 therapeutic classes.\\
The best model is estimated by Eq.(\ref{acemoglu}).\\
We introduced several regressors. These comprise supply-side determinants, technological opportunities, and age determinants. We draw some controls directly from the literature, comprising knowledge stock (see, e.g., \cite{cerda2007endogenous}, \cite{acemoglu2004market} among others) as measured by the number of papers referred to ATC category $i$. PubMed database has been consulted. Specifically, we collected the number of scientific works for a given ATC-3 in a given year through Mesh Terms. According to NIH, MeSH terms are official words or phrases selected to represent particular biomedical concepts. When labeling an article, indexers select terms only from the official MeSH list, never other spellings or variations. For deciding whether a paper referred or not to a specific ATC class, it has been first associated a Mesh Term to ATC category $i$ primarily exploiting the official synthetic description of ATC. If the latter did not produce any result or did not match evidence from literature, a double-check was made using level 3 indications as Mesh terms.\footnote{For instance, category C6B is described as "PULMONARY ARTERIAL HYPERTENSION (PAH) PRODUCTS." Due to the name's length and possible different abbreviations employed in the Mesh Terms list, Mesh Terms have been searched by looking at different specifications of the description such as "PAH PRODUCTS," "PULMONARY ARTERIAL HYPERTENSION PRODUCTS." If the latter did not produce any result or the results were not in line with findings in the literature, then Mesh indication at level 3 "PAH also searched terms." was also selected}. NCBI Mesh database allowed us to customize the searches. Since the number of papers showed an upward trend, the variable has been detrended through first differentiating its logarithm. \\
Another critical control drawn from literature is the share of generics. As noted in \cite{dubois2015market}, ease of entry and substantial financial incentives to use generics will reduce the expected profitability of the innovation. Hence, detecting the degree of penetration of generics within markets is vital, which might discourage firms from undertaking innovation. \\
Besides, as emphasized both in \cite{acemoglu2004market}, and \cite{dubois2015market}, a further source of declining margins of innovation is represented by the increasing number of young entrants within an ATC market. Pharmaceutical competition, in general, might undermine innovation productivity. It is, thus, imperative to measure and control for competition. \\
Apart from \cite{acemoglu2004market}, empirical literature does not model explicitly competition (see \cite{dubois2015market}). In the present work, we constructed two measures to control pharmaceutical competition. The first is the Herfindahl index ("hhi" hereafter), which measures firms' size in relation to the market. It is usually employed as an indicator of the amount of competition among firms within an ATC-3 market. The Herfindahl index's major benefit compared to other measures such as the concentration ratio gives larger firms more weight. The index can range from 0 to 1.0, moving from many tiny firms to a single monopolistic producer. 
The second measure controlling competition is the average age of firms within a market. It controls other aspects of competition compared to hhi. While hhi measures the "degree of monopoly" within an ATC, it cannot clarify the firms populating the market. However, the average age of firms mainly catches the presence of small biotechnology firms in the market. Such firms are known on one side to compete for innovation and on the other to have less financial resources in contrast with established companies (see, e.g., \cite{hall2010handbook} among others).
Since margins decline with the number of young entrants, we expect a negative sign of firms' average age.\\
Tab.\ref{secstage} presents the main results of the analysis. It is technically the second stage of the procedure described in the methodological section. Precisely, calling $z_{it2}$ the excluded instruments $(\text{recalls}_{it}, \text{recalls}_{it-1})$ \footnote{We included two instruments since, following \cite{hansen2008estimation}, instrumenting with more valid instruments leads to more accurate estimates}, the first stage estimation computes the residuals,$\ddot{u}_{it2}$, of a linear fixed-effect model whose dependent is market size. The second stage incorporates the residuals and estimates a fixed effect Poisson model. Please refer to steps 1. and 2. in the methodological section. \\
Differently from literature, in the present work, it is not necessary to construct $M_{it}$ based on demographic shifts since the innovative instrument, recalls, already purges market size from endogeneity. In the following, $M_{it}$ is simply the logarithm of collapsed sales at ATC-3 level, i.e., the product of the number of purchased drugs expressed in standard units to ensure comparability with their price. \\ Notice that a critical assumption of the model is that excluded exogenous, $R_{it}$ appearing within $z_{it}$, do not explicitly appear in the equation of Trials. For the more refined aggregation level at our disposal, ATC-3, it is plausible to assume that the average elasticity is the same across categories.

\begin{table}[H]\centering
\def\sym#1{\ifmmode^{#1}\else\(^{#1}\)\fi}
\captionsetup{justification=centering}
\caption{\footnotesize Impact of market size on innovation. Col.(1) employs a simple Poisson model \\ not considering fixed effects. Col.(2) is the main specification (fixed effect Poisson).\\ Col.(3) and Col.(4) add the lag of the dependent. Col.(5) eliminates all the controls \label{secstage}}
\scalebox{0.6}{ \vspace*{-0.01mm}\begin{tabular}{l*{5}{D{.}{.}{-1}}}
\toprule  & \multicolumn{1}{c}{\textbf{(1)}} & \multicolumn{1}{c}{\textbf{(2)}} & \multicolumn{1}{c}{\textbf{(3)}} & \multicolumn{1}{c}{\textbf{(4)}} & \multicolumn{1}{c}{\textbf{(5)}} \\
                         &\multicolumn{1}{c}{Trials}&\multicolumn{1}{c}{Trials} &\multicolumn{1}{c}{Trials}   &\multicolumn{1}{c}{$\textit{log Trials}$}&\multicolumn{1}{c}{Trials} \\
\midrule
$\textit{trials}_{t-1}$   & & & -0.00741&      0.0732\sym{*}  &                     \\
                         & & &  (0.0005)&    (0.0335)         &                     \\
Log sales                &      0.1378\sym{***} &      0.6362\sym{**} & 0.802\sym{**} &      0.1176\sym{***}&      0.8229\sym{**} \\ 
                                &    (0.0060)  &    (0.2149)  & (0.266) &    (0.0153)         &    (0.3174)           \\
residuals             &           &     -0.8018\sym{***} &  -0.862\sym{**} &                     &     -0.9711\sym{**} \\
                                 &        &    (0.2157)    & (0.269)&                     &    (0.3177)                 \\
$\frac{K_{t+1}}{P_{-1}}$        &     -0.5378\sym{***}  &     -0.0926      &-0.484\sym{***} &     -0.0504         &                     \\
                        &    (0.0847)    &    (0.0909)          & (0.147) &    (0.0914)         &                     \\
\textit{average age firm}&      0.2890\sym{***}    &     -0.1332\sym{***}     &-0.106\sym{*} &      0.0634\sym{**} &                     \\
                        &    (0.0139)     &    (0.0377)        & (0.0398) &    (0.0214)         &                     \\
\textit{average age firm}$^{2}$ &     -0.0038\sym{***}   &      0.0021\sym{***}  & 0.00178\sym{***}  &  -0.0008\sym{**} &                     \\
                        &    (0.0002)    &    (0.0005)    & (0.0005) &    (0.0003)         &                     \\
hhi                              &   0.2245\sym{***}  &     -0.3199 &  -0.145&      0.1106         &                     \\
                                  &    (0.0446)    &    (0.2903)   & (0.360)&    (0.1153)         &                     \\
\textit{share generics in ATC} &     -0.5571\sym{***}  &     -0.3168\sym{**} &-0.898\sym{***} &     -0.2036         &                     \\
                              &    (0.0404)    &    (0.1124)  & (0.143) &    (0.1068)         &                     \\
\textit{average age product}&   -0.0658\sym{***}    & -0.0592\sym{**}   &-0.0928\sym{**} &     -0.0564\sym{***}&                     \\
                        &    (0.0061).    &    (0.0190)      & (0.0323) &    (0.0143)         &                     \\
\textit{average age product}$^{2}$    &      0.0010\sym{***}  &      0.0011 &0.0100 &      0.0010\sym{*}  &                     \\
                         &    (0.0002)    &    (0.0010)     & (1.64)  &    (0.0004)         &                     \\
\textit{papers}           &      0.5608\sym{***}  &      0.1558\sym{*}  & 0.101   &     -0.0443         &                     \\
                                  &    (0.0728)     &    (0.0750)     & (0.0013)&    (0.1477)         &                     \\
\textit{papers}$^{2}$    &     -0.6672\sym{***} &     -0.0067      & -0.0929 &     -0.0083         &                     \\
                                  &    (0.1056)    &    (0.0838)      & (0.083)&    (0.0883)         &                     \\
\textit{\# firms}                &      0.0090\sym{***}      &      0.0008     &  -0.0032  &      0.0048\sym{**} &                     \\
                        &    (0.0007)     &    (0.0035)     & (0.0792) &    (0.0016)         &                     \\
\multicolumn{1}{l}{Year Dummies} & \multicolumn{1}{c}{Yes} & \multicolumn{1}{c}{Yes}  & \multicolumn{1}{c}{Yes} & \multicolumn{1}{c}{Yes} & \multicolumn{1}{c}{Yes}\\                                            \midrule  \multicolumn{1}{l}{Obs.} & \multicolumn{1}{c}{1664} & \multicolumn{1}{c}{1664} & \multicolumn{1}{c}{1664} & \multicolumn{1}{c}{1664} & \multicolumn{1}{c}{1872}  \\                                                                                        \multicolumn{1}{l}{Groups} & \multicolumn{1}{c}{208} & \multicolumn{1}{c}{208} & \multicolumn{1}{c}{208} & \multicolumn{1}{c}{208} & \multicolumn{1}{c}{208} \\
\bottomrule
\multicolumn{5}{l}{\footnotesize Standard errors in parentheses}\\
\multicolumn{5}{l}{\footnotesize \sym{*} \(p<0.05\), \sym{**} \(p<0.01\), \sym{***} \(p<0.001\)}\\
\end{tabular}}\begin{spacing}{0.1}\vspace{0.3cm} \footnotesize  {\fontsize{50}{60}\selectfont}{\fontsize{50}{60}\selectfont}{\fontsize{5}{6}\selectfont}{\Huge }{\begin{minipage}{12cm}\tiny Huber-White robust and clustered at ATC-3 level standard errors are in parentheses. (1) fits a simple Poisson with exogenous sales. (2) represents the main specification. The dependent variable is count of active trials in ATC $i$ at time $t$. Time interval is 10 years. The technique adopted for the estimation is the one of Wooldridge [2019], please refer to Section. \ref{methods}. (3) count model with lagged dependent among regressors following \cite{acemoglu2004market} (4) linear model with lagged dependent among regressors and exogenous size. Dependent is linearized. Both the presence of non-linearities and of endogeneity are ignored.(5) best model without controls   \end{minipage}}  \\ {\fontsize{50}{60}\selectfont}{\fontsize{5}{6}\selectfont}  \end{spacing}
\end{table}

Column (1) presents a simple Poisson model with exogenous market size exploring whether market size's positive effect is robust in the absence of fixed effects and endogeneity controls. Column (2) is our main specification, i.e., a fixed effect Poisson controlling for market size's endogeneity.\\
The coefficients of interest in Tab.\ref{secstage} are Log sales and residuals. The former represents the market size and the latter measuring endogeneity of market size. Specifically, a significant coefficient of residuals means a correlation between the error term (see the specification in Tab.\ref{secstage}, i.e., second stage regression) and functions of the error of the model of the market size (first stage). In other words, residuals control for co-movements of sales and unobservables related to the number of trials. Market size is hence "purged" from the alleged endogenous part. Endogeneity is tested with a Wald test on the coefficient $\rho$ of residuals. If $\rho$ is significantly different from zero, endogeneity is present. This latter instance occurs in our model as expected (Column (2)). In particular, fully robust standard errors detect a strong idiosyncratic endogeneity. The exploitation of Fixed Effects methodologies allows the unobserved heterogeneity to be correlated with all explanatory variables and the excluded exogenous recalls. The evidence is that even after allowing the market size to be correlated with the ATC heterogeneity, market size is not exogenous to idiosyncratic shocks.\\
The coefficient of market size is positive and significant in line with past works.  According to our estimate, a 10\% increase in market size leads to an increase of almost 6.3 \% of active trials. It turns out that also the magnitude conforms with literature.
Indeed, previous research generally finds elasticities to be approximately 0.5 consistently with our estimates. \\
Recent literature speculated on the possibility that, though clinical trials might respond elastically to market size, the proportion of them resulting in effective innovation might decline (see e.g.\cite{dubois2015market} among others). Hence authors might have overestimated the effect of market size on clinical trials since the latter should be computed only on the trials that effectively brought innovation. In the paper, we exploited active trials as a dependent, which partially solves the issue. We believe that active trials constitute the subset of promising trials in terms of innovative contribution. The estimated higher effect than the literature that adopts NMEs or NCEs as a dependent is well explained by the substantial costs for developing new pharmaceutical entities. Drug development is, in fact, quite expensive, the cost ranging between \$800 Million to \$2.5 Billion (see, e.g., \cite{medwatch}). Undertaking clinical trials is, instead, sensibly cheaper, amounting to an average of \$20 Million to \$40 Million (see \cite{martin2017much} as well as John Hopkins Bloomberg Health School, 2018). Thus it is reasonable to suppose that, ceteris paribus, a 10\% increase in market size stimulates more trials than NMEs or NCEs on average. Exceptions are still present (see \cite{acemoglu2004market}, \cite{duggan2010effect}, who estimated an higher elasticity than the one of the present work). \\
The coefficient of the average age of firms and its square is in line with past observations (see, e.g., \cite{huergo2004does} and \cite{balasubramanian2008firm} for specific studies on the topic). The effect evidences how the oldest firms tend to introduce less innovation than entrants in their early years. However, firms above intermediate ages \textit{appear almost as active in process innovations as entering firms and even more in product innovations} (\cite{huergo2004does}).\\
Moreover, innovation decreases with the share of generics within a market. Thus, the effect theorized in \cite{dubois2015market} of decreasing margins of innovation proportionally to the entrance of generics reveals to be correct (see also \cite{lanjouw2005patents}).\\
In line with \cite{acemoglu2004market} and \cite{rake2017determinants}, technological advancements as measured by detrended papers are positively related to innovation. It is indeed reasonable to suppose that more trials emerge in markets where scientific research is prolific. \\

The discrepancies in the magnitude of the coefficients between the main specification (Column (2)) and Column (1) of Tab.\ref{secstage} can be explained in several ways.  In Column (1) of Tab.\ref{secstage} correlation over time of units is not controlled. So it is assumed that units are independent over the cross-sectional dimension and over time dimension, which is quite a strong constriction in a longitudinal setting. The assumption means that the same individual (market) observed at two different times, $t_0$ and $t_1$, is considered independent from herself. In other words, individual (market) $i$ at time $t_0$ is another individual (market) than individual (market) $i$ at time $t_1$. The main implication of such presumption is that unobserved time-independent heterogeneities of individuals do not affect other individuals. However, we know that the same individual observed at two different times is considered "two distinct individuals." Thus, in the model of Column (1), it is ultimately assumed that unobserved shocks of an individual (market) $i$ at time $t$ do not influence individual (market) $i$ at time $t+k$. In other words, we are mixing between and within individual effects. Between effects are the effects obtained once the time component is averaged out from the variables. Between-effect settings exploit differences between units, which in our case are independent by definition (we take ATC-3 markets, see previous Sections), not taking into account time variations.  Therefore, the market size variance (time-demeaned) will be higher in a between-effect setting since it considers the average market size difference between independent ATC-3 markets.
Furthermore, given the opposite time trends of trials and market size (see Fig. \ref{tr_yearly}) in a between-effect setting, the between effects of market size on innovation will be deflated. Indeed, from a specific time on, the innovation trend decreases while the market size trend increases. However, since time variations are not controlled in a between-effect setting, the inverse proportionality of market size and innovation emerges. Mixing between and within individual effects will, hence, result in an overall lower coefficient of Column (1) compared to Column (2).\\
Ultimately, the downwardly biased coefficient of Column (1) suggests that the unobserved heterogeneity is negatively correlated to trials.\\
To provide an example, consider the possibility that an ATC experienced a sizeable positive shock (more trials) in 2010. For some reason, the mentioned shock is not modeled nor measured. All else being equal, the apparent fixed effect for that ATC in the period 2004-2013 will appear to be higher. However, from the literature, we know that the more the products available for treating a particular clinical condition, the lower the margins on each product (see \cite{bresnahan1991entry} among others). The unobserved positive shock for ATC $i^{th}$, therefore, would lower the margins of all competitor products in the same market, pushing down the sales for the same market. This negative correlation between the market size regressor and the error term deflates the estimate for market size. Vice versa, in Column (2), time dependency is controlled, and deflation is eliminated.  The coefficient of market size results, therefore, higher than in Column (1).
Column (1) does not control for the reverse causality of market size on innovation too. Not considering the reverse causality of market size contributes to upward biasing the market size's coefficient (see, e.g., \cite{acemoglu2004market}). There are, therefore, two contrasting effects: the upward effect due to the reverse causality endogeneity and the downward bias given by the unobserved heterogeneity endogeneity.
The two effects seem not to compensate, and negative heterogeneity bias prevails over reverse causality endogeneity bias.\\

\paragraph{Robustness checks} Col.(3)-(5) of Tab.\ref{secstage} investigate the robustness of the effect of market size on innovation. Three additional models are added to the preferred specification. Precisely, Column (3) reproduces the exercise of \cite{acemoglu2004market} to control for possibly varying over time technological flows (see below) by adding lagged trials among the regressors. Since the estimating
equation in Column (3) is nonlinear, we perform this instrumentation
strategy by adding the residuals of the first stage. Column (4) is the same as Column (3), where the dependent is log linearized, and residuals are ignored. Column (4) ignores both the presence of non-linearities and endogeneity.\\
Adding lags of the dependent is a valuable exercise. Indeed, following \cite{acemoglu2004market}, the primary threat to the identification strategy of innovation is represented by changes in the flow rate of innovation for every dollar spent for research on a drug (permanent differences in innovation are already dropped through the ATC fixed effects). Differences in the flow rate of innovation suggest that technological progress is scientifically more difficult in some lines than others. The parameter denoting innovation flow is part of the theoretical specification of innovation drawn from \cite{acemoglu2004market}. Following \cite{acemoglu2004market}, if the flow rate of innovation varies over time, it is also likely to be serially correlated. 
Adding lag of log innovation to the preferred specification is a simple way to check the importance of these concerns. The lagged trials are instrumented with their lags through a system GMM one-step procedure. The p-value of the Hansen test of overidentification of model in Column (4) is 0.175, falling mainly between the tolerance levels of 0.1 and 0.25 indicated in \cite{roodman2009xtabond2}. The Arellano-Bond test is investigated in Tab.\ref{tabab}.

\begin{table}[H]
\centering
 \begin{tabular}{||c| c| c||} 
 \hline
  & z-score & p-value   \\ [0.5ex] 
 \hline\hline
 \makecell{ Arellano-Bond test for AR(1) \\ in first differences:} & z = -10.47   & Pr > z =  0.000  \\ 
 \hline
 \makecell{ Arellano-Bond test for AR(2) \\ in first differences:} & z =   0.88  & Pr > z =  0.377  \\ 
 \hline
 \makecell{ Arellano-Bond test for AR(3) \\ in first differences:} & z =  -1.46  & Pr > z =  0.145  \\  
 \hline
 \makecell{ Arellano-Bond test for AR(4) \\ in first differences:} & z =  0.33   & Pr > z =  0.740  \\ 
 [1ex] 
 \hline
 \end{tabular}
 \caption[A-B test]{Arellano-Bond test for autocorrelation of first differenced residuals of GMM}
 \label{tabab}
\end{table}

When the idiosyncratic errors are independently and identically distributed (i.i.d.), the first-differenced errors are first-order serially correlated. So, as expected, the output above presents strong evidence against the null hypothesis of zero autocorrelation in the first-differenced errors at order 1. Yet, as suggested in Roodman (2009), \textit{"in the context of an Arellano-Bond GMM regression, which is run on first differences, AR(1) is to be expected, and therefore the Arellano-Bond AR(1) test result is usually ignored in that context"}. The output above presents, moreover no significant evidence of serial correlation in the first-differenced errors at order 2, 3 and 4. \\
In Column (4) market size is considered, again, exogenous, though fixed effects are controlled. The model in Column (4) is linear. In order to ensure comparability among models, trials have been transformed to a logarithmic scale. Column (4) is, in other words, an essential control since, though controlling for fixed effects, it ignores the presence of potential non-linearity (misspecification) and endogeneity, proposing the hypothesis of serial correlation. \\
Finally, Column (5) presents the model without any further control as estimated by the preferred specification's control function approach. The idea beyond Column (5) is to check whether not controlling for regressors compromises the main specification estimates.

The outcomes of Col.(3)-(5) of Tab.\ref{secstage} confirm the estimates of the main specification for what concerns the positive effect of market size on innovation. \\
Columns (3)-(5) in Tab.\ref{secstage} all display a positive effect of market size on innovation.\\
Specifically, Column (3) confirms the results of \cite{acemoglu2004market} finding no evidence of serial autocorrelation. In particular, the coefficient of lag trials is negative and non-significant as in \cite{acemoglu2004market}. Possible explanations are already in \cite{acemoglu2004market} and are, therefore, not discussed in the present work. 
In Column (4) of Tab.\ref{secstage}, the positive coefficient of lagged trials is significant at the 5\% tolerance level. This evidence is almost in line with \cite{acemoglu2004market} when no instrumentation is performed. \footnote{different from Column (1) where residuals of first-stage are included} Under this scenario the lagged dependent's coefficient turned out to be positive and not significant also in \cite{acemoglu2004market}. Market size is again strongly and positively related to innovation, with a coefficient having the lowest magnitude of the specifications analyzed until now. Indeed,  some of the variability might be caught by lagged dependent. Moreover, possible misspecification bias due to the not correction of nonlinearity might intervene. \\
Notice that the effect of the market size in Column (4) of Col. of Tab.\ref{secstage} display similarities to Col. (1) of the same table which does not control for endogeneity. Furthermore, the effect of market size is larger in the models correcting for endogeneity. Therefore, in general, the lack of control for temporal dependence may matter very little for estimation, as it is also consistent with the fact the autocorrelation coefficient is very weak. Otherwise, indeed, also the coefficient of size in  Columns (1) and (4) of Tab.\ref{secstage}, whose only dissimilarity relies on the control for temporal dependencies (Col.(4)), would have sensibly differed. Hence, it is reasonable to suppose that the lower magnitude of the coefficient of size in both Columns (1) of Tab.\ref{secstage}  and Column (4) of Tab.\ref{secstage} is primarily a consequence of considering market size as exogenous.
It is possible to provide further checks by controlling for possible overidentification of the instrumented lagged dependent variable. To do so, Tab.\ref{twstepgmm} column (1) reports the two-step robust system GMM estimates of Column (4) of Tab.\ref{secstage}, which, instead, performed a one-step system GMM.

\begin{table}[H]\centering
\def\sym#1{\ifmmode^{#1}\else\(^{#1}\)\fi}
\captionsetup{justification=centering}
\caption{\footnotesize Coefficients of market size and lag dependent when a two-step GMM is employed. \\
Col.(2) includes suspended and withdrawn trials in the dependent}\label{twstepgmm}
\scalebox{0.6}{ \vspace*{-0.01mm}\begin{tabular}{l*{2}{D{.}{.}{-1}}}
\toprule & \multicolumn{1}{c}{\textbf{(1)}} & \multicolumn{1}{c}{\textbf{(2)}} \\
                         &\multicolumn{1}{c}{\textit{log Trials}} &\multicolumn{1}{c}{\textit{log Trials}}\\ 
\midrule
$\textit{trials}_{t-1}$  &      0.0592   &      0.380\sym{*}      \\
                         &    (0.0426)  &      (0.189)       \\
Log sales                &      0.1208\sym{***} &      -0.533\sym{**}\\
&   (0.159)  & (0.179) \\
\text{Year Dummies} & \multicolumn{1}{c}{Yes}  & \multicolumn{1}{c}{Yes} \\     
\midrule  \multicolumn{1}{l}{Obs.} & \multicolumn{1}{c}{1664} & \multicolumn{1}{c}{1664} \\                                                                                        \multicolumn{1}{l}{Groups} & \multicolumn{1}{c}{208} & \multicolumn{1}{c}{208}  \\
\bottomrule
\multicolumn{2}{l}{\footnotesize Standard errors in parentheses}\\
\multicolumn{2}{l}{\footnotesize \sym{*} \(p<0.05\), \sym{**} \(p<0.01\), \sym{***} \(p<0.001\)}\\

\end{tabular}}\begin{spacing}{0.1}\vspace{0.3cm} \footnotesize  {\fontsize{50}{60}\selectfont}{\fontsize{50}{60}\selectfont}{\fontsize{5}{6}\selectfont}{\Huge }{\begin{minipage}{5cm}\tiny Huber-White robust and clustered at ATC-3 level standard errors are in parentheses. (1) is the two-step GMM version of Column (4) Tab.\ref{secstage}, which performed a system one-step GMM. Only the critical coefficients are included. (2) is equal to (1) where also suspended and withdrawn trials are included in the dependent. Only the critical coefficients are included. Both equations are linearized to enable a simple comparison with Column (4) Tab.\ref{secstage}. Thee same results apply if the count of trials is employed as dependent (see \cite{nutarelli})\end{minipage}}  \\ {\fontsize{50}{60}\selectfont}{\fontsize{5}{6}\selectfont}  \end{spacing}
\end{table}

The coefficient of the market size in Tab.\ref{twstepgmm} is higher compared to the one in Column (4) of Tab. \ref{secstage}. Furthermore, the lagged dependent variable is not significant in line with \cite{acemoglu2004market}. The same applies if the count of trials is employed as dependent (see \cite{nutarelli}).\\
The sign of the estimates of market size does not change with respect to the preferred model.\\
A final robustness check has been made by including all the trials, i.e., active ones and suspended and withdrawn. The number of classes employed is the same, though the number of trials increased by 0.57\% on the total. This is performed in Tab.\ref{twstepgmm} column (2).
As displayed, our estimation does not confirm the hypothesis in \cite{dubois2015market} showing a lower coefficient instead both in terms of magnitude and significance level. The hypothesis is that including non-active trials, which are less responsive to market size, biases estimate toward randomness. For instance, firms having all suspended trials are unaffected by price regulations reducing prices of treatments by governments. Simultaneously, increases in market size could be less effective on such companies, which already have sunk costs due to inactive trials. The presence of endogeneity is confirmed.\\

Further robustness checks have been performed by changing the market size's proxy to align to \cite{acemoglu2004market}, moving to another database to collect sales data (Evaluate sales are employed) and employing all the recalls at our disposal to instrument market size. In particular, Tab.\ref{fiststage} shows the outcomes of the analysis adopting Class II and Class I recalls as instrument for market size. Tab.\ref{pat_secs} measures market size through the number of patients within an ATC-3 in accordance with \cite{acemoglu2004market}. \\
Since the number of patients is highly correlated with sales and it is employed as a natural alternative to sales, we adopted recalls as an instrument for the number of patients.\\
The F-test amounts to 12 for the analysis with Evaluate and to 4 for the analysis with the number of patients.
\begin{table}[H]\centering
\def\sym#1{\ifmmode^{#1}\else\(^{#1}\)\fi}
\captionsetup{justification=centering}
\caption{\footnotesize Col.(1) and Col.(2) represent first and second stage results using all the recalls at our disposal. Data are aggregated at the ATC-3 level. Impact of market size on innovation using Evaluate database (Col.(3)) \and number of patients (Col.(4)) as proxy of market size \label{fiststage}}
\scalebox{0.6}{ \vspace*{-0.01mm}\begin{tabular}{l*{4}{D{.}{.}{-1}}}
\toprule & \multicolumn{1}{c}{\textbf{(First-stage all recalls)}}  & \multicolumn{1}{c}{\textbf{(Second-stage all recalls)}}  & \multicolumn{1}{c}{\textbf{(First-stage Evaluate)}} & \multicolumn{1}{c}{\textbf{(Second-stage Evaluate)}} \\
                         &\multicolumn{1}{c}{Log sales}&\multicolumn{1}{c}{Trials}&\multicolumn{1}{c}{Log sales}&\multicolumn{1}{c}{Trials}\\
\midrule
$\tilde{recalls}$         &     0.00199      &   & -0.260\sym{***}&   \\
                          &    (0.64)      &   &(0.0059) &          \\
$\tilde{recalls}_{t-1}$   &    -0.0223\sym{***} & & -0.0180&   \\
                          &     (-3.34)        &  & (0.0105)  &             \\
Log sales   &      &   0.580\sym{*} & &      0.710\sym{**} \\
                          &            &     (2.02)   &  &  (0.275)      \\
Residuals   &      &   -0.739\sym{*} & &     -0.724\sym{***}  \\
                          &            &     (-2.13)   & &    (0.275)     \\
$\frac{K_{t+1}}{P_{-1}}$ &      0.191\sym{*}       &  & -0.173 &    -0.147     \\
                          &           (3.07)           &  &  (0.154) &     (0.276)        \\
\textit{average age firm}
&          0.153           &       -0.129\sym{*}   & 0.0470 &      0.224\sym{***}       \\
                         
                          &      (1.67)                &    (-2.28) & (0.0678) &    (0.0045)         \\
\textit{average age firm}$^{2}$ &        -0.00195             &     0.00207\sym{**}  &-0.0004&     -0.0814\sym{***}       \\
                      &       (-1.53)              &    (2.87)  & (0.0008)&    (0.0316)        \\
hhi     &        1.238\sym{***}             &      -0.250\sym{***} &1.721\sym{***} &    -0.60\\
                 &          (4.79)           &    (-0.56)   & (0.419) &    (0.495)      \\
\textit{share generics in ATC}
&       -0.178                &     -0.326\sym{**} & -0.372&    0.00640        \\. 
                         
                          &    (-0.53)                 &    (-2.69)   & (0.328) &    (0.157)      \\
\textit{average age prod.} &        0.0539            &     -0.0559\sym{*}   &-0.0330&   -0.0732\sym{**}       \\
                      &       (0.92)              &    (-2.33)  & (0.0504)&    (0.0225)       \\
\textit{average age prod.}$^{2}$ &        -0.0037             &     0.0008  & 0.00138 &      0.0009      \\
                      &       (-1.72)              &    (0.63)    &(0.00206)& (0.0009)     \\
papers           &        -0.0267             &     0.153\sym{*} &-0.139&      0.265\sym{**}        \\
                  &          (-0.52)           &    (0.0507)   &(0.0803)&    (0.0917)       \\
\textit{\# firms}        
&          0.00729           &       0.00103  & -0.0106 &     0.0224\sym{***}        \\
                         &       (1.04)              &    (0.25)  & (0.0095) &    (0.0045)         \\
\multicolumn{1}{l}{Year Dummies} & \multicolumn{1}{c}{Yes} & \multicolumn{1}{c}{Yes}& \multicolumn{1}{c}{Yes}& \multicolumn{1}{c}{Yes} \\                                            \midrule  \multicolumn{1}{l}{Obs.}  & \multicolumn{1}{c}{1664} & \multicolumn{1}{c}{1664} & \multicolumn{1}{c}{1136}& \multicolumn{1}{c}{1056}  \\                                                                                        \multicolumn{1}{l}{Groups} & \multicolumn{1}{c}{208} & \multicolumn{1}{c}{208}& \multicolumn{1}{c}{142}& \multicolumn{1}{c}{132} \\
\bottomrule
\multicolumn{5}{l}{\footnotesize Standard errors in parentheses}\\
\multicolumn{5}{l}{\footnotesize \sym{*} \(p<0.05\), \sym{**} \(p<0.01\), \sym{***} \(p<0.001\)}\\
\end{tabular}}\begin{spacing}{0.1}\vspace{0.3cm} \footnotesize  {\fontsize{50}{60}\selectfont}{\fontsize{50}{60}\selectfont}{\fontsize{5}{6}\selectfont}{\Huge }{\begin{minipage}{13.5cm}\tiny Huber-White robust and clustered at ATC-3 level standard errors are in parentheses. The table shows the results when all recalls at our disposal are employed. First-stage results are shown Col. (1) while second stage results are in Col. (2). The level of aggregation is ATC-3 and the estimation method is the same as the one adopted in the main analysis. \end{minipage}}  \\ {\fontsize{50}{60}\selectfont}{\fontsize{5}{6}\selectfont}  \end{spacing}
\end{table}

Tab. \ref{fiststage} reports first and second stage of employing Class I and Class II recalls as instrument for market size in the first two columns. The remaining columns are devoted to the results obtained using Evaluate database to collect market sales. The results of the main analysis are confirmed in both exercises.\\
Employing all recalls, decreases both the magnitude and the significance of the coefficients of sales. Moreover only the lag of recalls is a good instrument at market level. These two effects are expected, since, including minor recalls may attenuate the drop in sales consequent to a recall. Indeed, within Class II are also comprehended temporary recalls (e.g. recalls due to a labeling error) which may both be not unexpected to the firm (most of them are voluntary) and, for this reason, taken into account by the management of the company. Losses in terms of sales are, therefore, well compensated. Furthermore, minor recalls are not publicized and cannot damage the image of the company or the market in which they happen.\\
Hence, adding minor recalls overtakes the strong and negative impact of current recalls and, as a consequence, affects the estimates of market size in the second stage. Since, however, Class II recalls, often regards minor but persistent issues\footnote{minor recalls often pertain the manufacturing of the product accessories. Their cause range from label mix-up, presence of particulate matter in certain lots to packaging issues. Though minor recalls do not threaten the health of patients directly, they are difficult to be corrected in the short-term by firms.}, a cumulative effect intervenes and lagged recalls remain a good instrument.\\
The outcomes of the main analysis remain robust when data on sales are collected from a different database. \\
Tab.\ref{pat_secs} reports the second stage results of the analysis with number of patients as a measure for market size. First stage results are in Appendix.\\
\begin{table}[H]\centering
\def\sym#1{\ifmmode^{#1}\else\(^{#1}\)\fi}
\captionsetup{justification=centering}
\caption{\footnotesize Impact of market size on innovation using number of patients as proxy of market size }\label{pat_secs}
\scalebox{0.6}{ \vspace*{-0.01mm}\begin{tabular}{l*{1}{D{.}{.}{-1}}}
\toprule &  \multicolumn{1}{c}{\textbf{(1)}} \\
                         &\multicolumn{1}{c}{Trials}\\
\midrule
Log patients                     &  3.274\sym{***} \\  
                                      &  (0.648) \\ 
residuals                &  -3.291\sym{***} \\ 
                                     &   (0.647)  \\    
$\frac{K_{t+1}}{P_{-1}}$         &     1.476\sym{***}   \\
                            &    (0.377)    \\       
\textit{average age firm}   &     0.589\sym{***} \\  
                           &    (0.154)    \\    
\textit{average age firm}$^{2}$   &  -0.00478\sym{**}  \\
                        &    (0.0016)    \\ 
hhi                              &     0.145  \\ 
                                     &    (0.260)   \\    
\textit{share generics in ATC}   &    -0.648\sym{**} \\ 
                                  &    (0.244)     \\       
\textit{average age product}    & -0.278\sym{***}   \\ 
                          &    (0.0514)     \\   
\textit{average age product}$^{2}$   &  0.0011\sym{***}  \\ 
                          &    (0.0022)       \\     
\textit{papers}           &      0.546\sym{***} \\ 
                                     &    (0.147)   \\     
$\textit{papers}^{2}$     &     0.193      \\    
                                    &    (0.114)   \\    
\textit{\# firms}                     &     0.0895\sym{***}  \\   
                             &    (0.0144)  \\ 
\text{Year Dummies} & \multicolumn{1}{c}{Yes} \\       \midrule  
\text{Obs.} & \multicolumn{1}{c}{1056} \\            
\text{Groups} & \multicolumn{1}{c}{132} \\
\bottomrule
\multicolumn{1}{l}{\footnotesize Standard errors in parentheses}\\
\multicolumn{1}{l}{\footnotesize \sym{*} \(p<0.05\), \sym{**} \(p<0.01\), \sym{***} \(p<0.001\)}
\end{tabular}} \begin{spacing}{0.1}\vspace{0.3cm} \footnotesize  {\fontsize{50}{60}\selectfont}{\fontsize{50}{60}\selectfont}{\fontsize{5}{6}\selectfont}{\Huge }{\begin{minipage}{9.6cm}\tiny Huber-White robust and clustered at ATC-3 level standard errors are in parentheses. (1) employs MEPS database and matches the ATC-3 present in our database. Market size is measured through the number of patients within ATC-3.  \end{minipage}}   {\fontsize{50}{60}\selectfont}{\fontsize{5}{6}\selectfont}  \end{spacing}
\end{table}

Adopting the number of patients as a proxy for market size confirms the first and second stage results compared to the principal specification. The outcomes, however, turn out to be weaker in terms of significance than our main specification. Recalls do not seem a strong instrument for the number of patients. On the one hand, a more significant number of patients within an ATC-3 class might increase the probability of an adverse event than in a scarcely populated ATC-3 class, increasing the probability of a major recall. On the other hand, however, there is no reason to believe that an adverse event would happen in a more populated class, which might refer to commonly employed medicines (and therefore well tested). Moreover, a recall in a class causes a decrease in the number of patients adopting pharmaceuticals in the questioned ATC-3 class, thus compensating the possible positive effect implied by a higher probability of adverse events. For what concerns the second stage, Tab.\ref{pat_secs} enforces the results found in \cite{acemoglu2004market} where a coefficient of market size between 3 and 4 was found. The significance of residuals confirms the presence of endogeneity.\\

\section{Conclusions} \label{concl}
Recent research has stressed the importance of market size in determining the innovation rate in the pharmaceutical industry. At the same time, after \cite{cerda2007endogenous} 's critique, instrumenting with demographical shifts remains a weak, though valid, strategy. Moreover, recent contributions have stressed the importance of modeling competition and technological opportunities adequately (see \cite{rake2017determinants}, \cite{dubois2015market}). For example, many scholars pointed to the importance of advances in molecular biology and related fields for the industry's technological opportunities
and innovative capabilities (\cite{rake2017determinants}). Finally, the literature lacks analyses at an aggregation level that easily allows drawing policy implications. For this reason, the present work employs ATC-3, the aggregation level used by Antitrust authorities.\\
The empirical estimates are conducted on a unique database integrated with additional sources. The variety of our sources enabled us to collect and adequately classify data on trials and drug recalls in ATC-3 categories. The methodology employed is innovative (\cite{lin2019testing}) and, differently from past techniques, permits controlling for both idiosyncratic endogeneity and heterogeneity endogeneity. The technique composes of two stages. A simple Wald test on the residuals' coefficient in the second stage allows verifying the presence of idiosyncratic endogeneity.\\
An innovative instrument, recalls, has been employed for the first time in literature. Recalls have been collected consulting various sources comprising FDA Enforcement reports, openFDA, and a database deriving from FOIA agreements with FDA. Recalls are representative. Major recalls have been selected to meet the criteria of sharpness, indirect effect on the dependent variable (innovation), and exogeneity. The first stage displayed a substantial and significant negative impact of recalls on market size, thus validating the instrument. To the best of our knowledge, the effort constitutes an empirical novelty in the literature that mainly focuses on optimal management of recalls and provides theoretical argumentation of recalls' negative impact at the firm level. Too few papers focused on the impact of drug recalls at the market level.\\ 
Data on clinical trials have been drawn from the Clinicaltrials.gov website from the pre-clinical phase to Phase IV. They have been integrated with data on INDs from a privately owned database maintained at IMT School for Advanced Studies. To overcome issues deriving from the potential more robust response of market size to trials as a whole rather than on essential trials (i.e., bringing most probably to an innovation), only activated trials have been selected. This exercise also provides a valid answer to the argumentation that the recall of a product might imply the suspension of drug trials within its same family. Indeed, suspended and withdrawn trials have been excluded from the analysis. Nonetheless, as a robustness check, estimates are computed also including the latter in the analysis. The effort confirms the presence of idiosyncratic endogeneity and the positive sign of the estimates. However, the magnitude and the significance level decrease.\\
Our preferred estimates align with literature displaying an increase in the innovation of 6.3\% after an increase in market size of 10\%. Most recent studies of \cite{dubois2015market} display a lower coefficient of 0.23\%. Authors specify how a comparison with other works exploiting different measures of innovation remains a difficult task. They further explain that their usage of global data rather than U.S. ones for the estimations might have led to less responsiveness.  \\
Our results are robust to several specifications. The coefficient of independent variables is in line with expectation as well as the scarce effect of lagged trials, already tested in \cite{acemoglu2004market}. Further checks confirm a positive and significant effect of market size on innovation even when fixed effects are not controlled, and the market size is considered exogenous. This latter verification partially validates (for what concerns the sign and the significance) the recent findings of \cite{rake2017determinants} who did not find evidence of reverse causality. However, the coefficient's magnitude decreases sensibly compared to the preferred specification, showing a significant bias.\\
Estimates remain robust even when no control is inserted in the analysis.\\
The work provides exciting policy implications for what concerns innovation's stimuli and sheds some light on the impact of recalls at the market level. Governments, in particular, should be aware wherever applying either tax or price policies in the pharmaceutical sector. As already mentioned, indeed, innovation constitutes an economic phenomenon. Companies innovate mainly to have a financial return. Aware of the positive relationship between market size and innovation, authorities and policymakers should not penalize economic players too much. To guarantee citizens' future welfare, they should promote research and invest in new technologies smartly managing generics' competition. \\
Recalls, moreover, have not just an impact at the firm level but also the market level. Specifically, they provoke adverse shocks on markets, thus affecting economic stability and welfare. Authorities should therefore apply more stringent rules to avoid severe recalls. At the same time, they should consider that an intensification of Class II and Class III recalls due to the presence of more players might be physiological.\\
Future research might employ more up-to-date data in order to include also recalls of compounders and repackaging firms. 
\paragraph{Acknowledgements:} \footnotesize We are grateful to Crisis Lab. for the PHID database. We thank Prof. Jeff Wooldridge for the kind support in the application and interpretation of his innovative methodology. We would also like to thank the FDA for providing data on recalls and clearing doubt on recalls' timing and procedures. We are grateful to Evaluate data for having allowed very up-to-date and detailed checks on clinical trials data. Furthermore, we are thankful to Springer AdInsight for having enabled a detailed classification at ATC 3 level of rare clinical trials. Finally, we thank Young Economists of Tuscan Institutions (YETI), the AXES unit of the IMT Lucca for Advanced Studies  and the ASSA meeting participants for the valuable comments on the work. 

\clearpage

\newpage
\appendix
\section*{\\Appendices}
\renewcommand{\thesection}{\Alph{section}}
\renewcommand{\thesubsection}{\Alph{section}.\arabic{subsection}}
\section{Literature review table}
The following table summarizes the literature's findings on the relationship between market size and innovation in the pharmaceutical industry. In particular, the focus is on relevant works coming after \cite{acemoglu2004market}. The reason for such a choice is that \cite{acemoglu2004market} represents a milestone in investigating the relation between market size and innovation in pharmaceuticals. It overcomes issues emerging in previous studies (such as, and above all, the one of endogeneity) and is taken as a reference point by authors willing to further dig into such a literature stream.\\
Furthermore, the literature review reports the relationship between market size and innovation in Pharmaceutical Industry only. Indeed, different industries have different definitions of recalls.

\begin{table}[H]
\setlength{\tabcolsep}{6pt} 
\renewcommand{\arraystretch}{2.5}
\caption{The table reports relevant papers after \cite{acemoglu2004market}; NME stands for New Molecular Entities; NDA stands for New Drug Approval.}
\centering
\scalebox{0.45}{
\begin{tabular}{||c| c| c| c| c| c| c||} 
 \hline
 Paper & \makecell{Data and \\ sample} & Unit of observation & \makecell{Measurement of \\ innovation} & estimation method & \makecell{report estimate \\ of size} &\makecell{proxy market\\ size} \\ [0.5ex] 
 \hline
 \cite{acemoglu2004market} & \makecell{US; March CPS, 1965–2000;
 \\ March CPS, 1965–2000; FDA; OECD} & report number of units & NME $^b$ & QML & > 0 $^4$ & demographic measures  \\
 \cite{cerda2007endogenous} & \makecell{US; FOIA request; RND process$^1$;
 \\ \& U.S statistical abstract$^2$; 1968-1997} & 15 drug categories $^1$  & NME & FE, GLS, IV, Tobit & >0 $^1$ & demographic measures \\
 \cite{rake2017determinants} & \makecell{U.S.; RND; FDA; OECD; \\ ClinicalTrials.gov$^3$; 1974-2008} & disease & \makecell{NDA; NME; Phase II \\ and Phase III trials} & QMLE (Poisson, 1995) & 0.3444 (NME); 0.3521 (NDA) & demographic measures\\
 \cite{dubois2015market} & 14 countries $^5$; 1997-2007;IMS, WHO & \makecell{chemical entity; \\ dummies for ATC-1 and ATC-2}  & NCE (elasticity) $^b$ & \makecell{OLS,2SLS,CF \\approach (Wooldr.,2002)} & \makecell{0.23 (average across\\ ATC classes)} & deaths and GDP $^5$ \\ 
 \cite{blume2013market} & \makecell{US; 1998-2010; Pharmaprojects $^6$; \\ MEPS; OECD; NIH} & 49 therapeutic classes & R\&D $^6$ & Negative Bin.; Poisson & 0.26; 0.41; 0.51 $^7$ & demographic shifts $^5$ $^c$
\\
 \hline
\end{tabular}} 
\begin{spacing} {0.1} \vspace{0.3cm}  {\fontsize{50}{60}\selectfont}{\fontsize{50}{60}\selectfont}{\fontsize{5}{6}\selectfont}{\Huge } {\begin{minipage}{15cm}\tiny 
$^1$RND process of the pharmaceutical sector (gov. funds); > 0 means that the exogenous increase in market size is initially associated with approximately 0.08 more drugs introduced in the market. These new drugs reduce the
mortality rates of individuals aged 65 and older by 0.8 percent. This decrease in mortality rate leads to increases in market size (more demand), producing an additional increase of drugs equal to 0.096\\
$^2$(population data for market size)\\
$^3$ Both Cerda and Rake consulted the 19th edition of the Drug Information Handbook published by Lexi-Comp and the American Pharmaceutical Association (Lacy et al., 2010). This handbook is comparable to a pharmaceutical dictionary, providing a list of drugs' active ingredients, the medical conditions the drug is used for, and further information such as adverse effects. The work takes into account only those medical conditions which can be found on the FDA-approved label. Hence, unlabeled and investigational uses are not present. For the period 1974 to 2008, FDA approved 599 unique NMEs and 1,665 unique NDAs. These approvals refer to the 208 diseases or medical indications analyzed in this study. However, an NME or NDA may be used as therapy for several medical indications. In this case, an NME or NDA is counted as innovation for all the medical indications for which it is approved\\
$^4$ The estimates suggest that a 1 percent increase in the potential market size for a drug category leads to a 6 percent increase
in the total number of new drugs entering the U. S. market. \\
$^5$  Data come from IMS (Intercontinental Marketing Services) and include all product sales in 14 countries $^a$ (Australia, Brazil, Canada, China, France, Germany, Italy, Japan, Mexico, Korea, Spain, Turkey, United Kingdom, USA). Dubois et al. have data on the ATC-4 (they report 607 different classes), the main active ingredient of the drug (they report 6216 different active ingredients), the name of the firm producing the drug, whether it has been licensed, the patent start date, and the format of the drug (the work reports 471 different formats). Products in the same ATC-4 by definition have the same indication and mechanism of action. The authors do not consider OTC drugs. Quantities are given in standard units, one standard unit corresponding to the smallest typical dose of a product form, as defined by IMS Health.\\
$^6$ Pharmaprojects trend data "snapshot"; (focus on R\&D): focus only on one instance of innovation as explained in \cite{hall2010handbook}. Authors specify the adoption of clinical trials (from pre-clinical Phase to Phase III) not taken from ClinicalTrials.gov (see below) \\
$^7$ For a drug class with average Medicare market share (41\%, in 2004–2005), Duggan and Scott Morton's result translates to an 11\% increase in revenues following Medicare Part D. Our Phase I estimates correspond, for a drug class with average Medicare market share, to a 26\% increase for 2004–2005, a 33\% increase post-implementation in 2006–2007, and a lagged 51\% increase in 2008–2010. These estimates imply an elasticity of Phase I clinical trials of 2.4 to 4.7 compared to the market size, bracketing Acemoglu and Linn's estimated elasticity of 3.5 for approved new molecular entities (NMEs). However, when considering all clinical trials combined—including Phase III trials for supplemental indications the estimated elasticity of clinical trials with respect to market size is somewhat lower than Acemoglu and Linn's estimated elasticity of 6 for all new drug approvals, but certainly still more prominent than the Dubois et al. (2011) estimate of about 0.25. Summary results: "The results indicate that the increase in outpatient prescription drug coverage provided through Medicare Part D has had a significant impact on pharmaceutical R\&D "\\~\\
Critiques: \\
$^a$  \cite{blume2013market} states that several of the countries chosen regulate prescription drug prices, and regulations may change rapidly over time. Thus, given the lower expected profit per consumer and greater uncertainty about future profits and prices, firms’ R\&D decisions are likely to be less responsive to a unit change in expected revenues for all these countries combined versus the same unit change in the U.S. market (Sood et al., 2009).\\
$^b$\cite{blume2013market}:  they measured firms’ innovative activities via clinical trials, whereas Dubois et al. (2011) and Acemoglu and Linn (2004) evaluate the responsiveness of approved and marketed drugs to changes in market size \\
$^c$  \cite{dubois2015market}: the authors recognize to \cite{blume2013market} the fact of having exploited an innovative measure of Market Share (policy change in Medicare Part D) \\
List of controls: \\
\cite{acemoglu2004market} Potential Supply-Side Determinants of Innovation (changes in scientific incentives); Proxies for pre-existing time trends across sectors; lag dependent var; life-years lost; public funding; pre-existing trends; major category trends; health insurance market size; (see page 1077-1080 for further details on variables)\\
\cite{cerda2007endogenous}: Gov. expenditure (Medicare and social security); Gov. research efforts (grants on research); year dummies; some demographic information such as prevalence rates of disease $i$ on males (fraction of males/white/married attending hospital due to $i$), blacks, whites, and married individuals as well as the average age of individuals affected by disease $i$. \\
\cite{rake2017determinants} The empirical analysis draws upon the literature concerning the “demand-pull” versus “technology-push” debate and takes into account demand- and supply-side factors as the explanatory variables for pharmaceutical innovation. Regressors used comprise knowledge stock (consisting of the scientific publications (Pub$_{it}$) related to medical indication $i$ and published in year $t$ (BioPharmInsight database); Regulatory stringency (average time between the submission of a new drug approval to the FDA and its final approval); pre-sample mean of new pharmaceuticals;  mortality rate
per medical indication in 1983 to account for differences in the pre-sample prevalence of medical indication; pre-sample technological opportunities are constructed as the average annual growth rate of the knowledge stock from 1979 to 1983.\\
\cite{blume2013market} prescription drugs; funding grants for each disease class  

\end{minipage}}  \\ {\fontsize{50}{60}\selectfont}{\fontsize{5}{6}\selectfont}  \end{spacing}

\label{litrev}
\end{table}

\section{Time to event analysis}
\begin{figure}[H]
  \centering
\includegraphics[width=7cm]{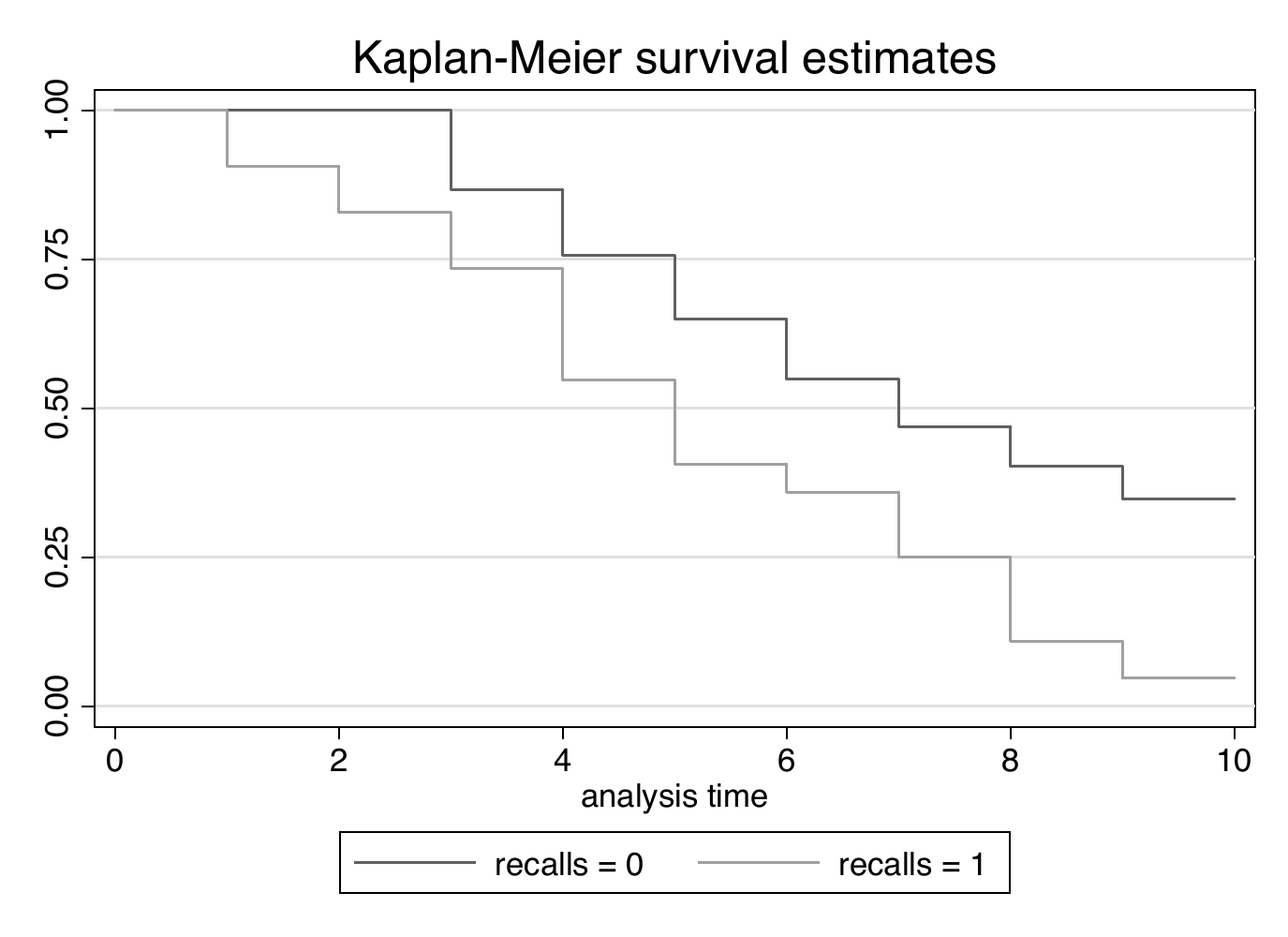}\newline
   \caption{\footnotesize Kaplan-Meyer time to event analysis. The x-axis represents the number of years until death. Recalled products (recalls =1) are already scaled down at 2 years of survival time with respect to not recalled products (recalls =0). The median survival time for recalled medicines is about 4 years, while the one for not-recalled medicines is about 7 years. The survival function of recalled products persists in its falling below the survival function of not recalled drugs. This means that recalls affect sales for a long period of time. In other words, within the market of the recalled product there will be a lack of potential sales left by the recalled products. Missed sales are hence not a temporary event, demonstrating the length of the lack that should be covered to fill the gap provoked by the product's recall.}
  \label{km_sur}
\end{figure}

\section{Abnormal Values (firm and product levels)}

\begin{figure}[H]
  \centering
\includegraphics[width=7cm]{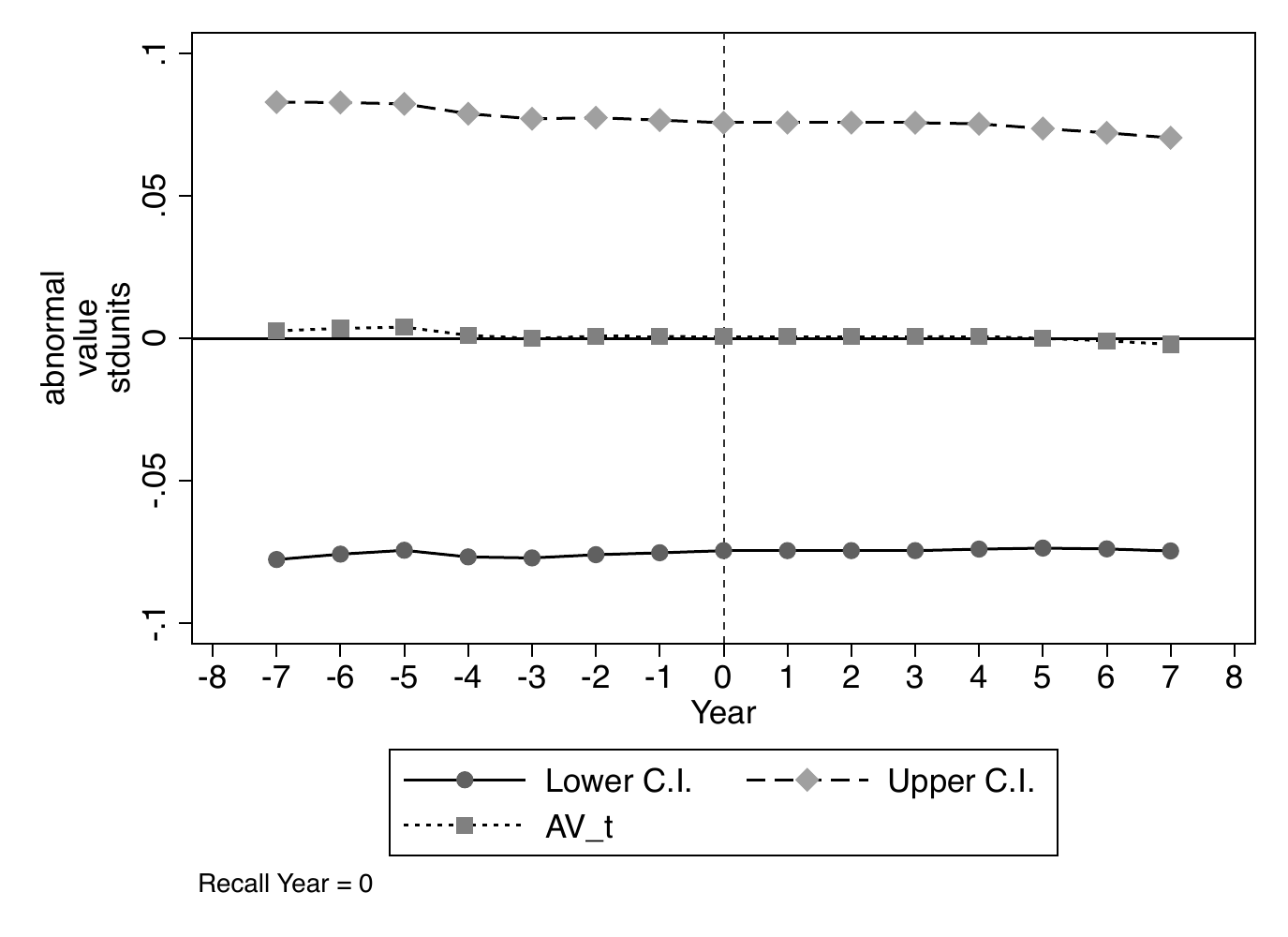}\newline
  \caption{\footnotesize Effect of recalls on market sales once firms having undergone a major recall are cancelled out. The absence of any effect (i.e. increases of sales due to recalls of competitors) at recall time for products other than the ones of the recalled firm witnesses the absence of compensations both at time 0 or soon after the recall.} 
  \label{sales_nowithd}
\end{figure}

\newpage
The following figures represent abnormal values at firm and product level for different typologies of recall (according to their gravity).
\begin{figure}[H]
  \centering
   \begin{adjustbox}{minipage=\linewidth,scale=0.8}
\hfil\hfil\includegraphics[width=7cm]{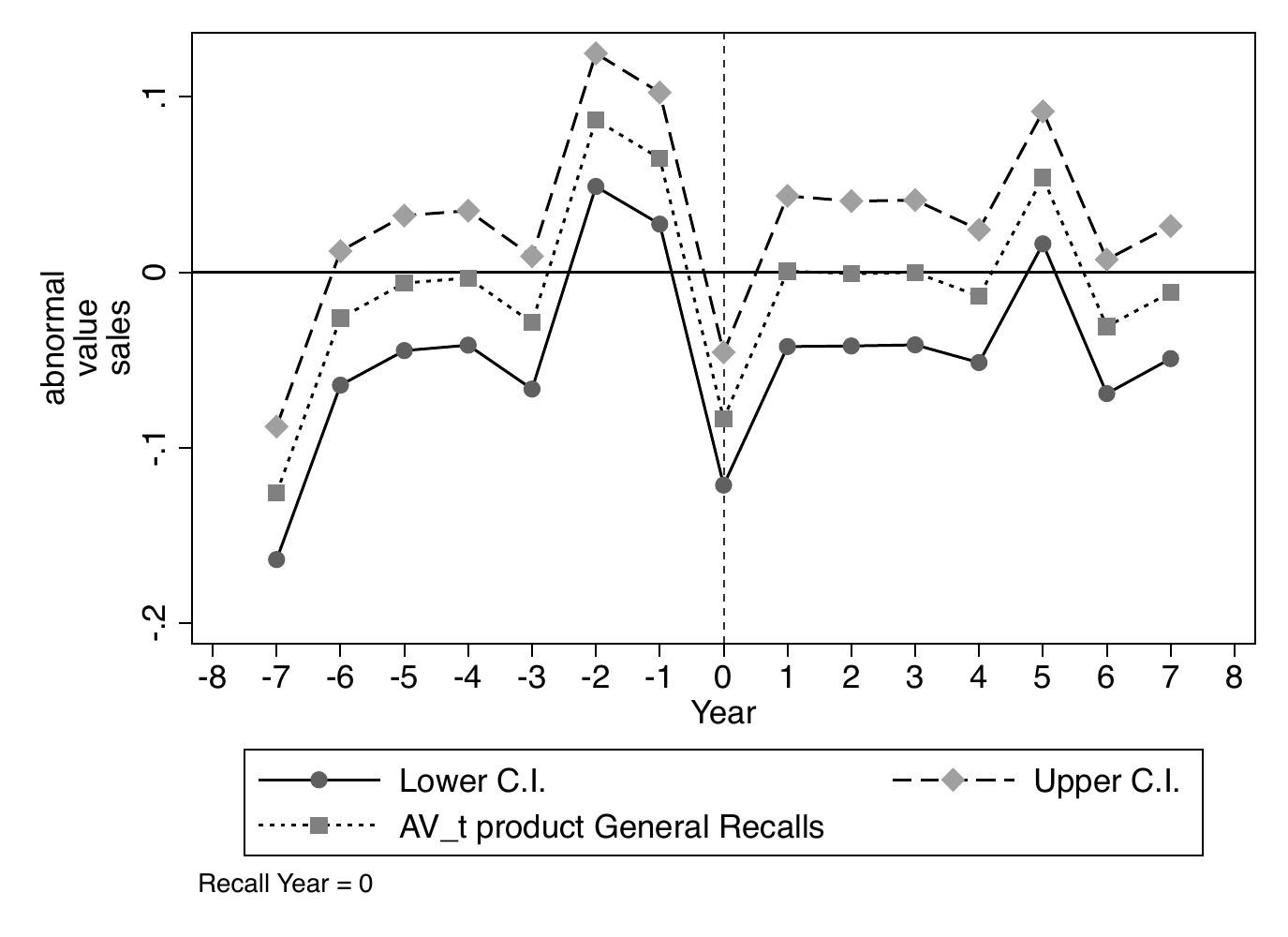}\newline
  \null\hfil\hfil\makebox[5cm]{Product: general recalls sales}\newline
  \vfil
  \hfil\hfil\includegraphics[width=7cm]{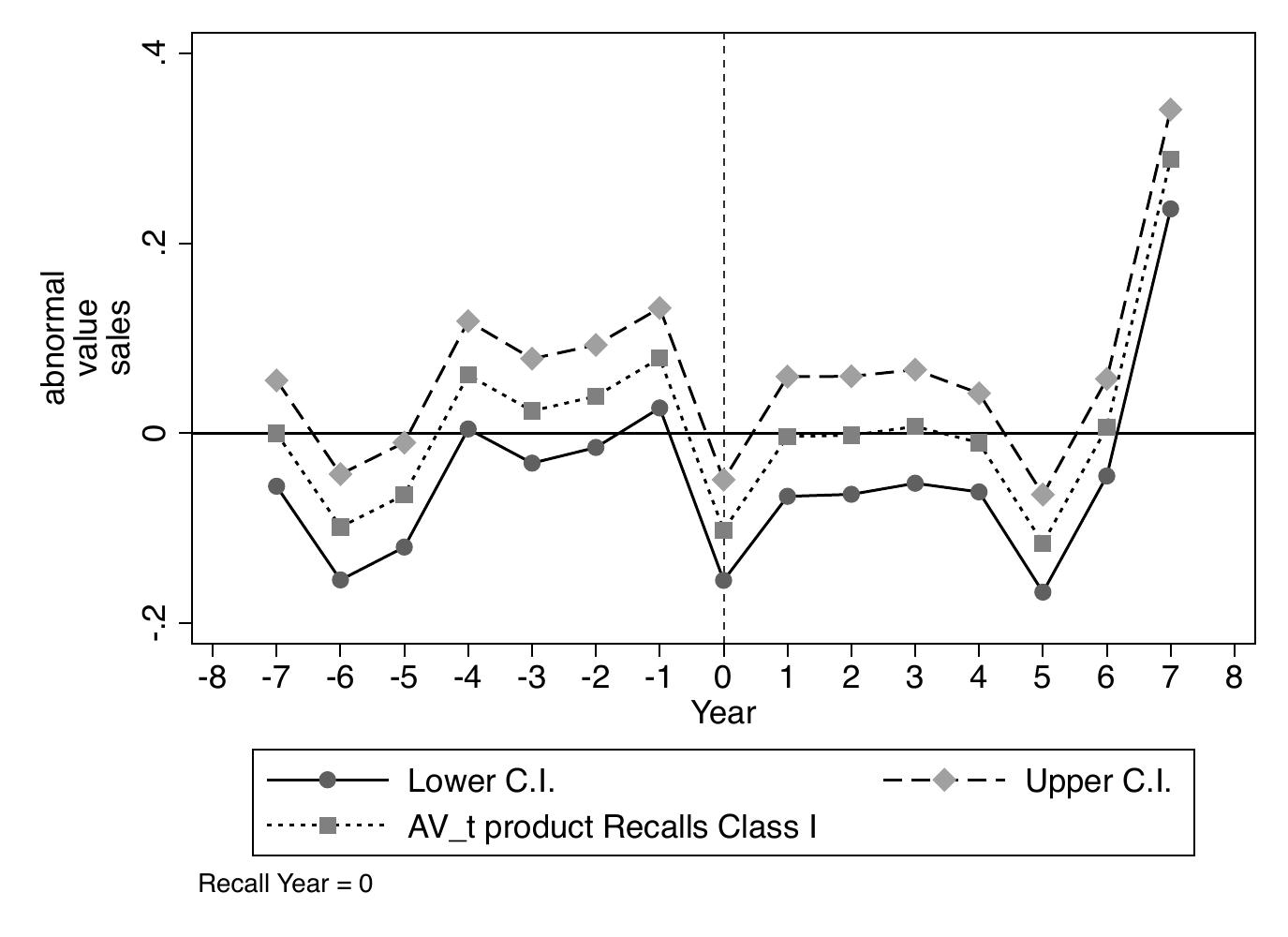}\hfil\hfil
    \includegraphics[width=7cm]{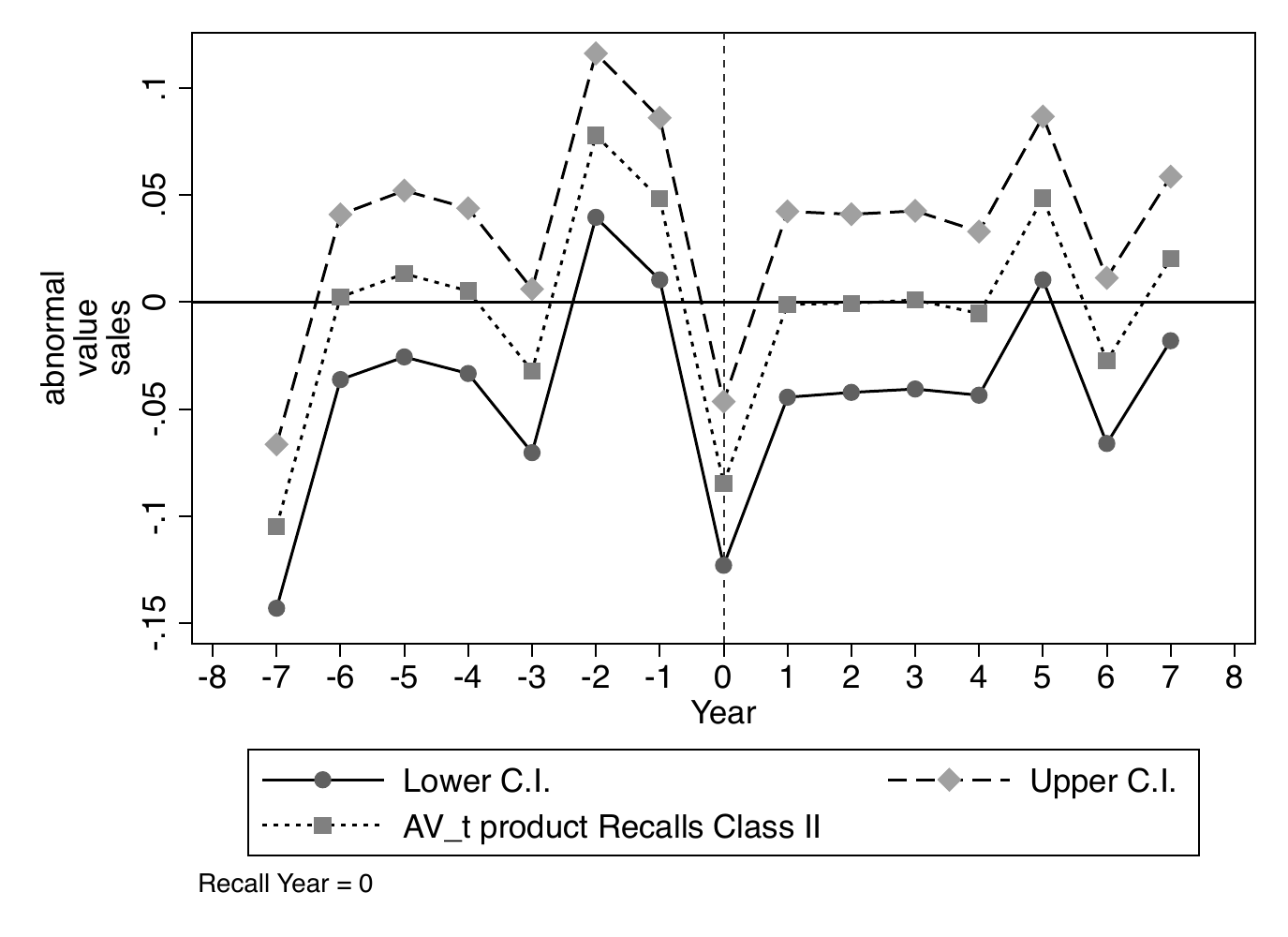}\newline
  \null\hfil\hfil\makebox[5cm]{Product: Class I recalls sales}
    \hfil\hfil\makebox[5cm]{Product: Class II recalls sales}
  \caption{Abnormal values for product aggregation. Years are normalized. Year 0 represents the year of recall. The three scenarios include the path of sales before and after the recall year, using three different definitions of recalls: Class I recalls, general recalls and Class II recalls. As shown in the pictures, sales at product level drop at recall year.}
  \label{product_sales}
  \end{adjustbox}
\end{figure}
\vspace{0.5cm}
\begin{figure}[H]
  \centering
   \begin{adjustbox}{minipage=\linewidth,scale=0.8}
  \hfil\hfil\includegraphics[width=7cm]{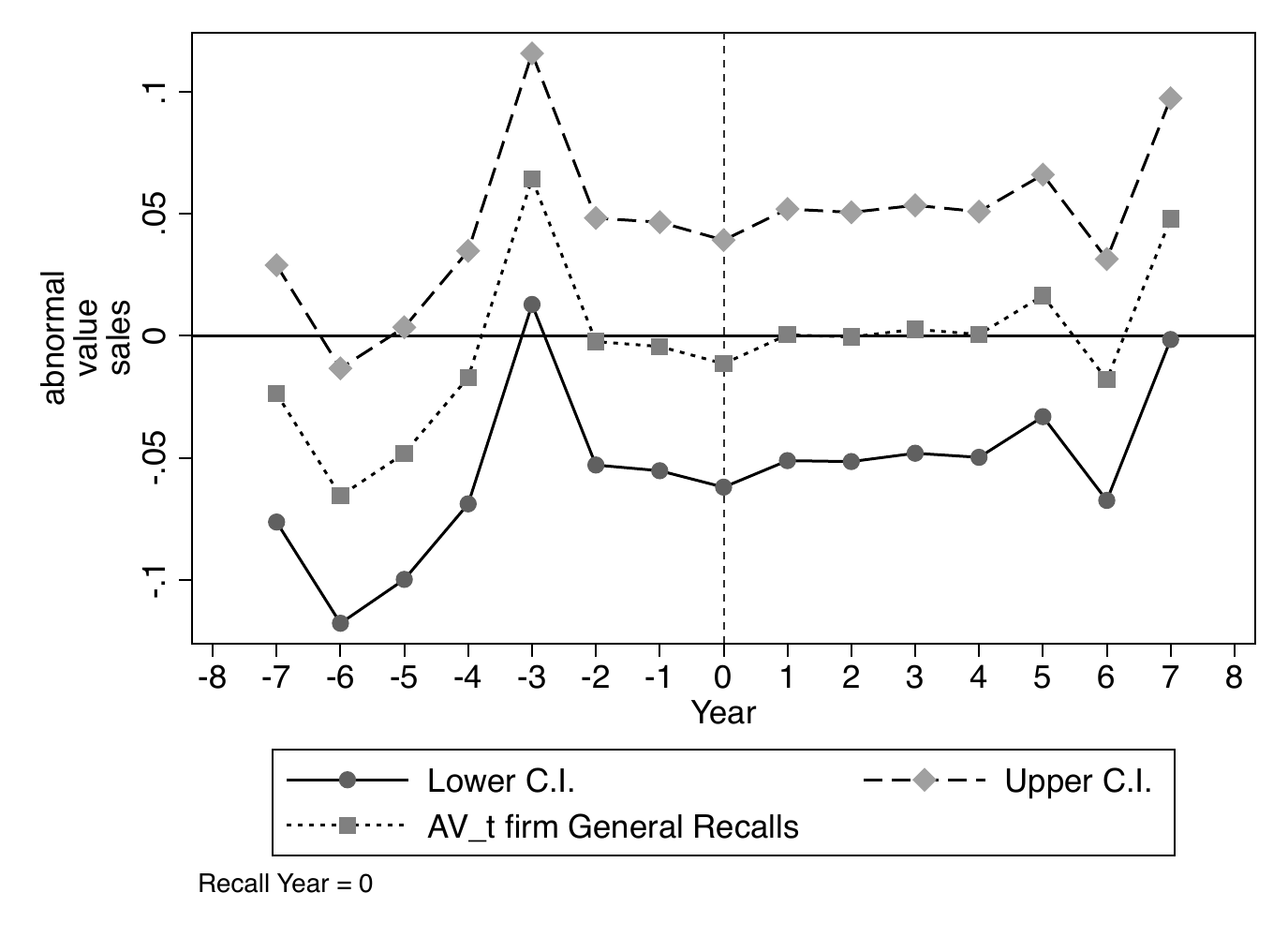}\hfil\hfil
    \includegraphics[width=7cm]{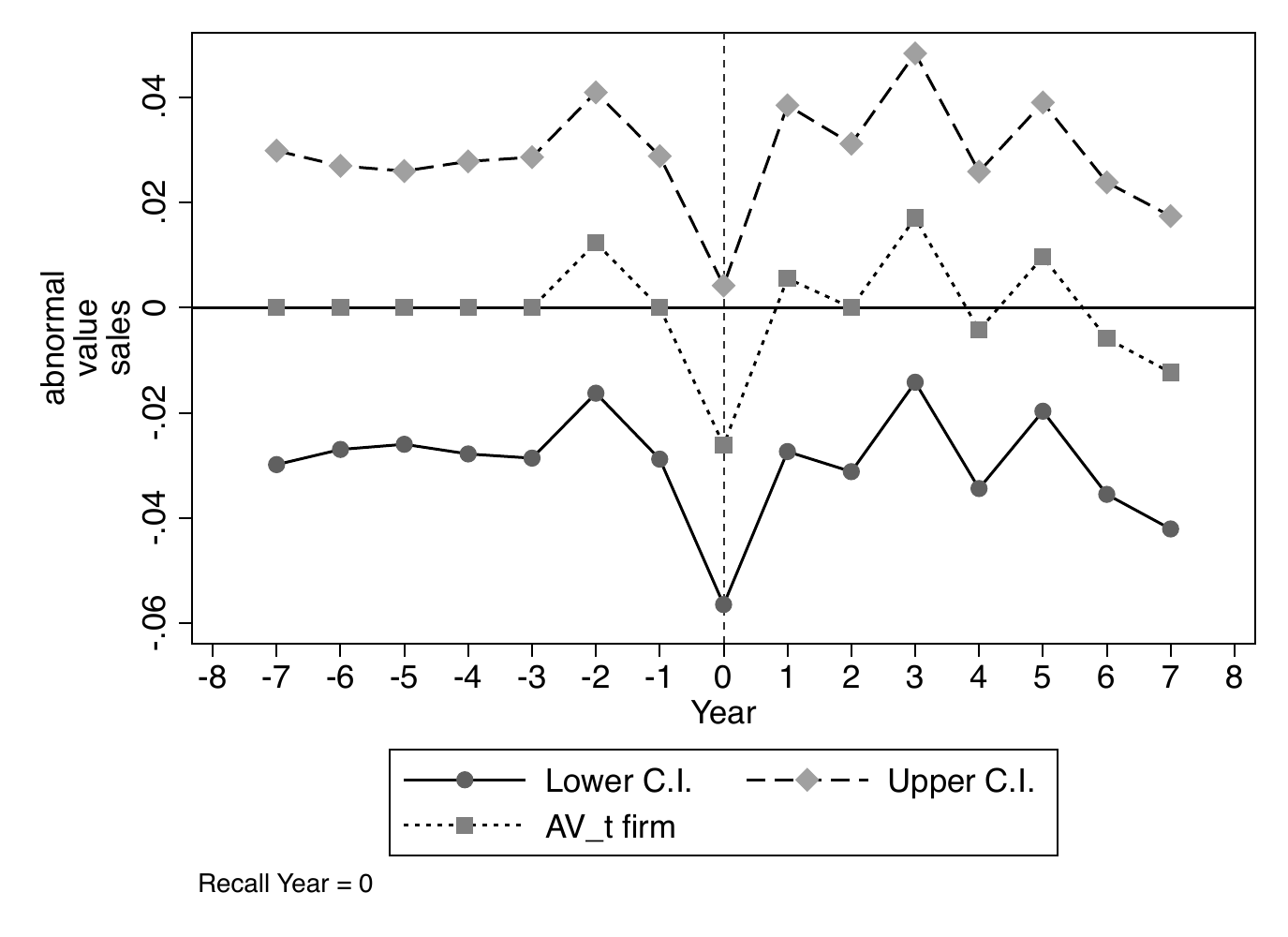}\newline
   \null\hfil\hfil\makebox[5cm]{Firm: general recalls sales}
    \hfil\hfil\makebox[5cm]{Firm: Maj.recalls sales}
  \vfil
  
  \hfil\hfil\includegraphics[width=7cm]{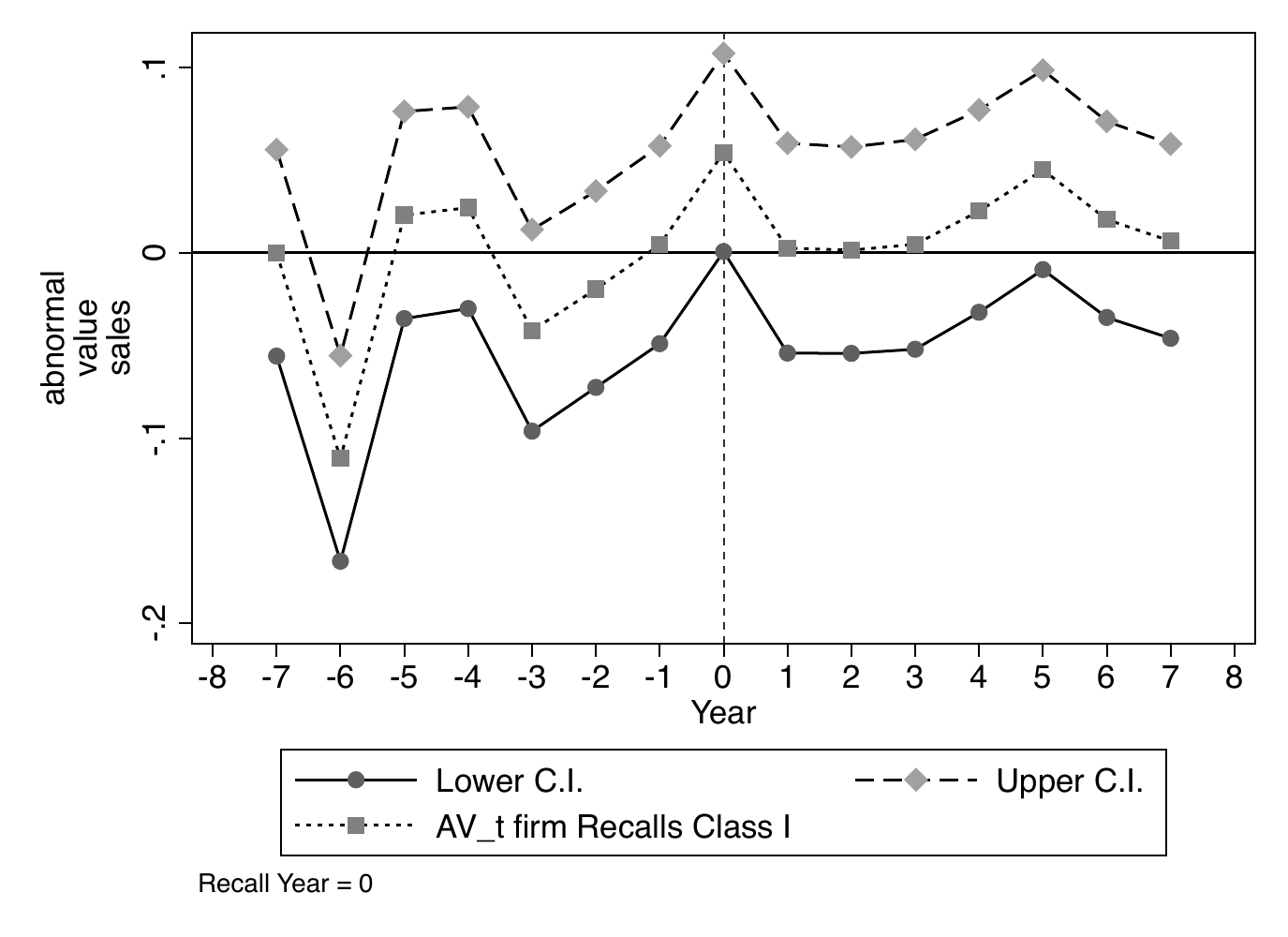}\hfil\hfil
    \includegraphics[width=7cm]{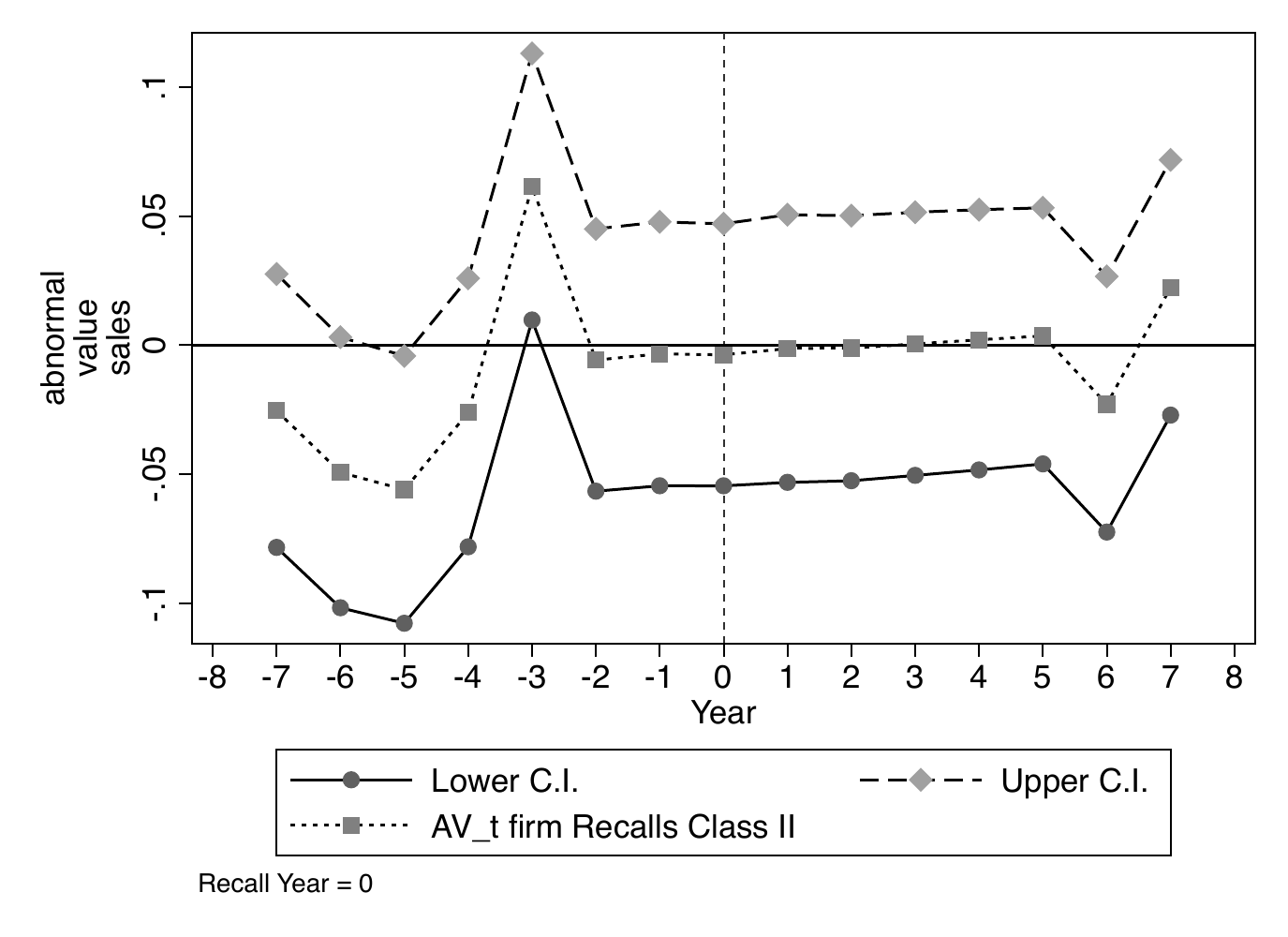}\newline
  \null\hfil\hfil\makebox[5cm]{Firm: Class I recalls sales}
    \hfil\hfil\makebox[5cm]{Firm: Class II recalls sales}
  \label{fig:firm_sales}
  \caption{Abnormal values at firm aggregation. Years are normalized. Year 0 represents the year of recall. The four scenarios include the path of sales before and after the recall year, using four different definitions of recalls: major recalls, Class I recalls, general recalls and Class II recalls. A part from major recalls, catching firms unaware, the other types of recalls do not affect firms sales. This might be due to compensation of sales within firms.}
  \end{adjustbox}
\end{figure}

The effect of recalls is evident for all aggregations but firm level, where the effect is not evident (future development).Possible hypotheses are detailed in the main text.

\section{First stage rob. checks}
In this section are displayed the significant coefficients of the first stage employing the number of patients as measure for market size.
\begin{table}[H]\centering
\def\sym#1{\ifmmode^{#1}\else\(^{#1}\)\fi}
\caption{\footnotesize First stage of the robustness check using the number of patients as measure of market size \label{firststagepresc}}
\scalebox{0.6}{ \vspace*{-0.01mm}\begin{tabular}{l*{1}{D{.}{.}{-1}}}
\toprule & \multicolumn{1}{c}{\textbf{(1)}} \\
&\multicolumn{1}{c}{$\textit{\# patients}$} \\ 
\midrule
$\tilde{recalls} $     &  0.0519   \\ 
                        &    (0.0333)  \\ 
$\tilde{recalls}_{t-1}$     &  -0.262\sym{*} \\ 
 &  (0.149) \\
\multicolumn{1}{l}{Year Dummies} & \multicolumn{1}{c}{Yes} \\
\midrule  \multicolumn{1}{l}{Obs.} & \multicolumn{1}{c}{1056} \\  
\multicolumn{1}{l}{Groups}  & \multicolumn{1}{c}{132} \\
\bottomrule
\multicolumn{2}{l}{\footnotesize Standard errors in parentheses}\\
\multicolumn{2}{l}{\footnotesize \sym{*} \(p<0.05\), \sym{**} \(p<0.01\), \sym{***} \(p<0.001\)}\\
\end{tabular}}\begin{spacing}{0.1}\vspace{0.3cm} \footnotesize  {\fontsize{50}{60}\selectfont}{\fontsize{50}{60}\selectfont}{\fontsize{5}{6}\selectfont}{\Huge }{\begin{minipage}{5cm}\tiny Huber-White robust and clustered at ATC-3 level standard errors are in parentheses. (1) first stage results when MEPS database is employed and market size is measured with the number of patients. Second stage results are in Tab. \ref{pat_secs}. \end{minipage}}  \\ {\fontsize{50}{60}\selectfont}{\fontsize{5}{6}\selectfont}  \end{spacing}
\end{table}

\clearpage
\newpage
\bibliographystyle{plain}
\bibliography{nog}
\nocite{alsharkas2014firm}
\nocite{jovanovic2020product}
\nocite{ball2018recalls}
\nocite{bala2017pharmaceutical}
\nocite{dimasi2003price} 
\nocite{dubois2015market} 
\nocite{hall2010handbook} 
\nocite{hashi2013impact} 
\nocite{kolluru2017empirical} 
\nocite{luan2013downsizing}
\nocite{pammolli2011productivity}
\nocite{symeonidis1996innovation}
\nocite{stock2002firm}
\nocite{enwiki:1053053987}
\nocite{vujovic2021case}
\nocite{enwiki:1045065285} 
\nocite{provost}
\nocite{hwkk}
\nocite{markham2020lurbinectedin}
\nocite{cheng2008antitrust}
\nocite{vaishnav2011product}

\end{document}